\documentclass[USenglish,oneside,twocolumn]{article}

\usepackage[utf8]{inputenc}
\usepackage[big]{dgruyter_NEW}
\usepackage[margin=0.1cm,font=small]{caption}
\usepackage{makecell}
\usepackage{xspace}
\usepackage{xcolor}
\usepackage{float}
\usepackage{soul}
\usepackage[font=footnotesize]{subcaption}

\usepackage[colorinlistoftodos,prependcaption,textsize=footnotesize,color=yellow]{todonotes} 
\makeatletter
\usepackage{regexpatch}
\xpatchcmd{\@todo}{\setkeys{todonotes}{#1}}{\setkeys{todonotes}{inline,#1}}{}{}
\makeatother

\newcolumntype{C}[1]{>{\centering\let\newline\\\arraybackslash\hspace{0pt}}m{#1}}
\newcolumntype{L}[1]{>{\raggedright\let\newline\\\arraybackslash\hspace{0pt}}m{#1}}

\newcommand{\first}{\textsl{first-party-collection-use}\xspace}

\newcommand{\thirds}{\textsl{third-party}\xspace}
\newcommand{\firsts}{\textsl{first-party}\xspace}
\newcommand{\pinfo}{\textsl{info-type}\xspace}
\newcommand{\retention}{\textsl{data-retention}\xspace}
\newcommand{\category}{\textit{category\xspace}}
\newcommand{\purpose}{\textit{purpose\xspace}}
\newcommand{\polisis}{\textsl{Polisis}\xspace}

\newcommand{\sone}{\textit{s1}\xspace}
\newcommand{\stwo}{\textit{s2}\xspace}
\newcommand{\sthree}{\textit{s3}\xspace}

\newcommand{\unspec}{\textsl{unspecified}\xspace}

\newcommand{\pregdpr}{\text{pre-GDPR}\xspace}
\newcommand{\postgdpr}{\text{post-GDPR}\xspace}

\cclogo{\includegraphics{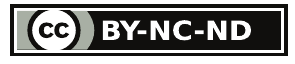}}

\usepackage[moderate]{savetrees}

\makeatletter
\g@addto@macro\normalsize{%
\setlength\abovedisplayskip{2pt}
\setlength\belowdisplayskip{2pt}
\setlength\abovedisplayshortskip{3pt}
\setlength\belowdisplayshortskip{3pt}
}
\makeatother

\makeatletter
\renewcommand*{\@seccntformat}[1]{\csname the#1\endcsname\hspace{0.3cm}}
\makeatother
\begin{document}

\author[1]{Thomas Linden}
\author[2]{Rishabh Khandelwal}
  \author[3]{Hamza Harkous}
  \author*[4]{Kassem Fawaz}

  \affil[1]{University of Wisconsin, E-mail: tlinden2@wisc.edu }
  \affil[2]{University of Wisconsin, E-mail: rkhandelwal3@wisc.edu }

  \affil[3]{École Polytechnique Fédérale de Lausanne, E-mail: hamza.harkous@gmail.com}

  \affil[4]{University of Wisconsin, E-mail: kfawaz@wisc.edu}

%   \affil[4]{Affil, E-mail: email@email.edu}
  \title{\huge The Privacy Policy Landscape After the GDPR}

  \runningtitle{The Privacy Policy Landscape After the GDPR}

  \begin{abstract}
{
The EU General Data Protection Regulation (GDPR) is one of the most demanding and comprehensive privacy regulations of all time. A year after it went into effect, we study its impact on the landscape of privacy policies online. We conduct the first longitudinal, in-depth, and at-scale assessment of privacy policies before and after the GDPR. We gauge the complete consumption cycle of these policies, from the first user impressions until the compliance assessment. 
We create a diverse corpus of two sets of 6,278 unique English-language privacy policies from inside and outside the EU, covering their \pregdpr and the \postgdpr versions.
The results of our tests and analyses suggest that the GDPR has been a catalyst for a major overhaul of the privacy policies inside and outside the EU. This overhaul of the policies, manifesting in extensive textual changes, especially for the EU-based websites, comes at mixed benefits to the users.\\ 
While the privacy policies have become considerably longer, our user study with 470 participants on Amazon MTurk indicates a significant improvement in the visual representation of privacy policies from the users' perspective for the EU websites. 
We further develop a new workflow for the automated assessment of requirements in privacy policies.
Using this workflow, we show that privacy policies cover more data practices and are more consistent with seven compliance requirements post the GDPR.
We also assess how transparent the organizations are with their privacy practices by performing specificity analysis. In this analysis, we find evidence for positive changes triggered by the GDPR, with the specificity level improving on average. Still, we find the landscape of privacy policies to be in a transitional phase; many policies still do not meet several key GDPR requirements or their improved coverage comes with reduced specificity. 
\vspace{-2\baselineskip}
}
\end{abstract}

\maketitle

\section{Introduction}

For more than two decades since the emergence of the World Wide Web, the ``Notice and Choice'' framework has been the governing practice for the disclosure of online privacy practices. This framework follows a market-based approach of voluntarily disclosing the privacy practices and meeting the fair information practices~\cite{marotta2016self}. The EU's recent General Data Protection Regulation (GDPR) promises to change this privacy landscape drastically. As the most sweeping privacy regulation so far, the GDPR requires information processors, across all industries, to be transparent and informative about their privacy practices.

\subsubsection*{Research Question}
   Researchers have conducted comparative studies around the changes of privacy policies through time, particularly in light of previous privacy regulations (e.g., HIPAA\footnote{The Health Information and Portability Accountability Act of 1996.} and GLBA\footnote{The Gramm-Leach-Bliley Act for the financial industry of 1999.}) ~\cite{anton2007hipaa,anton2004financial,adkinson2002privacy,milne2002using}. Interestingly, the outcomes of these studies have been consistent: (1) the percentage of websites with privacy policies has been growing, (2) the detail-level and descriptiveness of policies have increased, and (3) the readability and clarity of policies have suffered.

The GDPR aims to address shortcomings of previous regulations by going further than any prior privacy regulation.
One of its distinguishing features is that non-complying entities can face hefty fines, the maximum of 20 million Euros or 4\% of the total worldwide annual revenue. Companies and service providers raced to change their privacy notices by May $25^{th}$, 2018 to comply with the new regulations~\cite{degeling2018we}. With the avalanche of updated privacy notices that users had to accommodate, a natural question follows: 

\textit{What is the impact of the GDPR on the landscape of online privacy policies?}

Researchers have recently started looking into this question by evaluating companies' behavior in light of the GDPR. Their approaches, however, are limited to a  small number of websites (at most 14)~\cite{contissa2018claudette,tesfay2018privacyguide}. Concurrent to our work, Degeling et al.~\cite{degeling2018we}, performed the first large-scale study focused on the evolution of the cookie consent notices, which have been hugely reshaped by the GDPR (with 6,579 EU websites). They also touched upon the growth of privacy policies, finding that the percentage of sites with privacy policies has grown by 4.9\%. 

\subsubsection*{Methodology and Findings}
Previous studies have not provided a comprehensive answer to our research question. In this paper, we answer this question by presenting the first on-scale, longitudinal study of privacy policies' content in the context of the GDPR. We develop an automated pipeline for the collection and analysis of 6,278 unique English privacy policies by comparing \pregdpr and \postgdpr versions. These policies cover the privacy practices of websites from different topics and regions. We approach the problem by studying the change induced on the entire experience of the users interacting with privacy policies. We break this experience into five stages:

\textbf{A. Presentation (Sec.~\ref{sec:presentation}).}
To quantify the progress in privacy policies' \textit{presentation}, we gauge the change in user perception of their interfaces via a user study involving 470 participants on Amazon Mechanical Turk. Our results find a positive change in the attractiveness and clarity of EU-based policies; however, we find that outside the EU, the visual experience of policies is not significantly different.

\textbf{B. Text-Features (Sec.~\ref{sec:readability}).}
We study the change in the policies' using high-level syntactic text features. Our analysis shows that both \textit{EU} and \textit{Global} privacy policies have become significantly longer, with (+35\%, +25\%) more words and (+33\%, +22\%) more sentences on average respectively. However, we do not observe a major change in the metrics around the sentence structure.

We devise an approach, inspired by goal-driven requirements engineering~\cite{van2001goal}, to evaluate coverage, compliance, and specificity in the privacy policies. While previous longitudinal studies either relied on manual investigations or heuristics-based search queries~\cite{adkinson2002privacy,anton2004financial,degeling2018we}, we build on the recent trend of automated semantic analysis of policies. We develop a total of 24 advanced, in-depth queries that allow us to assess the evolution of content among the set of studied policies.  We conduct this analysis by building on top of the \polisis framework (Sec.~\ref{sec:reqs}), a recent system developed for the automated analysis of privacy policies~\cite{harkous2018polisis}. Following this approach, we perform a deeper level of semantic analysis that overcomes the limitations of keyword-based approaches.

\textbf{C. Coverage (Sec.~\ref{sec:cat_coverage}).}
We evaluate the policies' coverage of high-level privacy practices. For both \textit{EU} and \textit{Global} policies, we find a significant improvement in the policies' coverage of topics highly relevant to the GDPR such as data retention (52\% improvement for \textit{EU}, 31\% improvement for \textit{Global}), handling special audiences (16\% improvement for \textit{EU}, 12\% improvement for \textit{Global}), and user access (21\% improvement for \textit{EU}, 13\% improvement for \textit{Global}).

\textbf{D. Compliance (Sec.~\ref{sec:compliance}).}
We design seven queries that codify several of the GDPR \textit{compliance} metrics, namely those provided by the UK Information Commissioner (ICO). The GDPR's effect is evident; for both of our analyzed datasets, we find a positive trend in complying with the GDPR's clauses: substantially more companies improved (15.1\% for \textit{EU} policies and 10.66\% for \textit{Global} policies, on average) on these metrics compared to those that worsened (10.3\% for \textit{EU} policies and 7.12\% for \textit{Global} policies, on average).

\textbf{E. Specificity (Sec.~\ref{sec:specificity}).}
Finally, we design eight queries capturing how specific policies are in describing their data practices. Our results show that providers are paying special attention to make the users aware of the specific data collection/sharing practices; 25.22\% of \textit{EU} policies and 19.4\% of \textit{Global} policies are more specific in describing their data practice. Other providers, however, are attempting to cover more practices in their policies at the expense of specificity; 22.7\% of \textit{EU} policies and 17.8\% of \textit{Global} policies are less specific than before.

Building on the above, we draw a final set of takeaways across both the time and geographical dimensions (Sec.~\ref{sec:conclusion}).

\section{GDPR Background}
\label{sec:background}

As the most comprehensive privacy regulation to date, the General Data Protection Regulation (Regulation (EU) 2016/679), passed on April 14, 2016 and enforced on May 25, 2018, is the European Union's approach to online privacy. The GDPR defines four entities: data subjects, data controllers, data processors, and third parties. The data subjects are the users of the information systems from which data is collected. The data controller is typically the service provider (e.g., website or mobile app) with a vested interest in receiving and processing the user data. A data controller might employ a processor to process the data on its behalf. Finally, the data controller might authorize a third party (e.g., analytics agency) to process some of the user's data.  

Chapter III of the GDPR describes the rights of the data subjects; the first (Article 12) is the right to be informed about the service provider's privacy practices ``in a concise, transparent, intelligible and easily accessible form, using clear and plain language.'' The service provider has to communicate its practices regarding data collection and sharing (Articles 13 and 14) as well as the rights of users associated with data collection and processing (Articles 15-22).

Under the GDPR, the service provider has to inform the user about the contact information of the controller, the purposes for data collection, the recipients of shared data, the retention period and the types of data collected. Furthermore, the service provider has to notify the users about updates to its privacy practices promptly. Articles 13 and 14 make a distinction between data collected directly from the user or obtained indirectly. The service providers have to inform the users about the source and type of information when obtained indirectly.
Articles 15-22 list the rights of users regarding data collection and processing. These rights include the right of access, rights of rectification and erasure, right to the restriction of processing, right to data portability and right to object. In this paper, we focus on the requirements set in Articles 12-22 of the GDPR and study how privacy policies evolved in meeting these requirements.

\section{Creation of The Policies Dataset}
\label{sec:methodology}

We assembled two datasets of privacy policies, one to represent EU-based privacy policies, and a second representing policies from around the world. Hereafter, we will refer to the former as the \textit{EU} set and the latter as the \textit{Global} set. For each policy, we selected two snapshots between January 2016 and May 2019: \pregdpr and \postgdpr. We define the \pregdpr snapshot as the last stable version policy before the enforcement of the GDPR. The \postgdpr is the most recent version of the policy.  We restrict our analysis to privacy policies in the English Language. 

\subsection{Methodology}
\begin{figure}[t!]
    \centering
    \includegraphics[width=\linewidth]{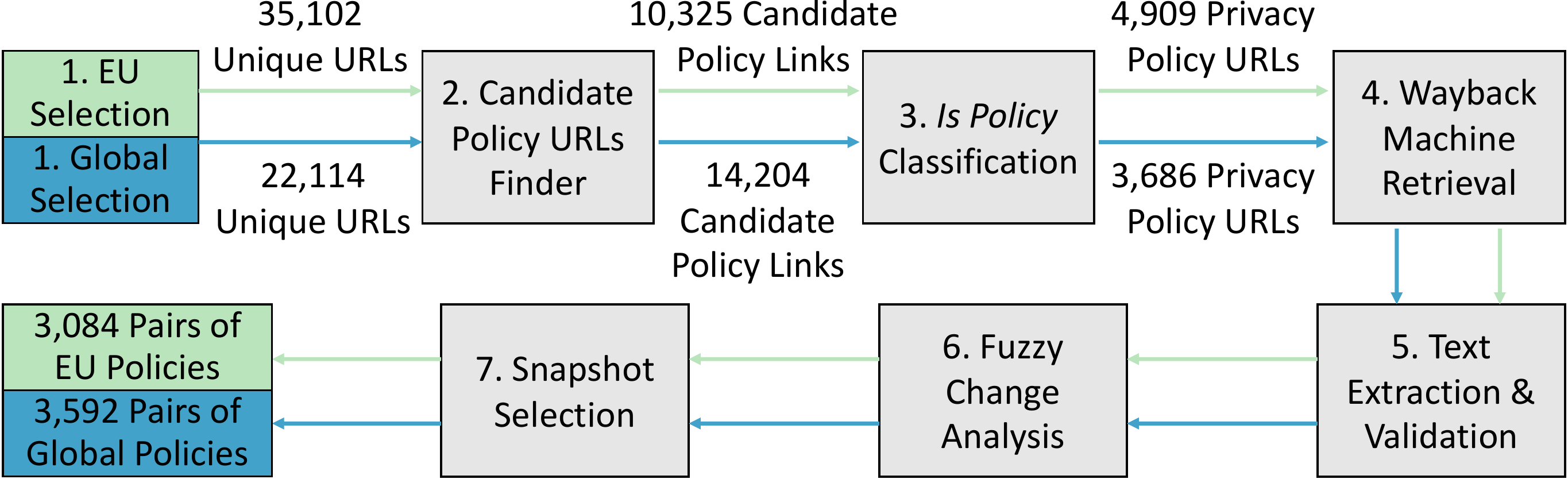}
    \caption{Our methodology of retrieving the policy URLs.}
    \label{fig:policy_url_retriever}
\end{figure}

In the following, we describe our methodology to create a corpus of privacy policies as highlighted in Fig.~\ref{fig:policy_url_retriever}.

\paragraph*{Website Selection}

To select the set of websites for the \textit{EU} set, we used the Alexa TopSites API \footnote{\url{https://www.alexa.com/topsites/}} to query the top 3,000 domains for each of the 28 EU member-states. This step resulted in 82,389 URLs, as some of the smaller states, such as Malta, returned fewer results (2,394). Of the extracted domains, we found 35,102 unique instances. For the \textit{Global} set, our methodology aimed at selecting websites that exhibit a topical and geographical mix. We used the Alexa Website Information Service\footnote{\url{https://aws.amazon.com/awis/}} to obtain the top links, ranked by popularity, in each of the 16 Alexa categories, spanning different topics (e.g., \textit{Adult, Arts, Business, Regional}). We amended these categories by considering 9 subcategories of the \textit{Regional} category (e.g., \textit{North America, Middle East, Europe}). For each of the 25 categories, we considered the top 1,000 visited websites. This step resulted in a set of 25,000 URLs, of which we counted 22,114 unique URLs. We note that the starting \textit{EU} set was significantly larger by design, as our restriction to English policies excludes a significant portion of the candidate \textit{EU} domains.

\paragraph*{Policy Finder}

We automatically crawled the home page of each of the URLs identified in the previous stage. We crawled the HTML using the Selenium framework\footnote{https://www.seleniumhq.org/} and a headless Chrome Browser. We identified a set of candidate privacy policy links on the home page based on regular expressions (e.g., the presence of words like \textit{privacy, statement, notice,} or \textit{policy} in the URL or the title). This stage resulted in candidate sets of 10,325 \textit{EU} pages and 14,204 \textit{Global} pages.
In a lot of cases, initially distinct URLs share the same privacy policy link due to the same company owning multiple websites (e.g., YouTube owned by Google or Xbox owned by Microsoft).

\paragraph*{Is-Policy Classification}

 We developed a custom \textsc{Is-Policy} classifier to decide whether each of the candidate pages belongs to a valid English-language privacy policy. The \textsc{Is-Policy} classifier consists of two modules. The first is a language detection module, using \textit{langid.py}~\cite{lui2012langid} that labels non-English websites as \textit{invalid}. The second module is a one-layer Convolutional Neural Network (CNN) that we developed to output the probability that the input text belongs to a privacy policy (based on the classifier by Kim~\cite{Kim14}). The accuracy of the classifier is 99.09\% accuracy on the testing set. The details of the classifier's architecture and training ar in Appendix~\ref{sec:pol_classifier}.

 The \textsc{Is-Policy} classifier assigned a \textit{valid} label (i.e., an English privacy policy) to 4,909 \textit{EU} URLs, as well as 3,686 \textit{Global} URLs. Besides language, additional reasons for a candidate URL's rejection included the domain's ``robots.txt'' policy, the lack of any privacy policy at the candidate link, and the privacy policy embedded within the longer terms of service. We considered these rejected cases to be non-suitable for our further automated analysis, prioritizing high precision over high recall.

\paragraph*{Wayback Machine Retrieval}

We used the Wayback Machine\footnote{\url{https://archive.org/help/wayback_api.php}} from the Internet Archive to collect a series of raw HTML snapshots for each privacy policy from the last stage. We used the python library \textit{Waybackpack}\footnote{\url{https://pypi.org/project/waybackpack/}} to request every unique archived instance from Jan. 2016 to May 2019, with a monthly resolution (\textit{i.e.} at most one instance was returned per month).

\paragraph*{Text Extraction and Validation}

For each downloaded HTML file of each policy, we used Boilerpipe~\cite{kohlschutter2010boilerplate} to get the cleaned HTML of the webpage without the unnecessary components (e.g., headers and footers). We extracted the body text using the library \textit{Beautiful Soup}\footnote{\url{https://www.crummy.com/software/BeautifulSoup/}}. After these sanitization steps, a portion of the snapshots was reduced to very short text bodies and was removed from further processing. We performed further manual verification of a random set of the remaining snapshots to ensure that the false positives have been removed.

\begin{figure}
\captionsetup[subfigure]{justification=centering}
     \centering
     \begin{subfigure}[b]{0.49\columnwidth}
         \includegraphics[width=\textwidth]{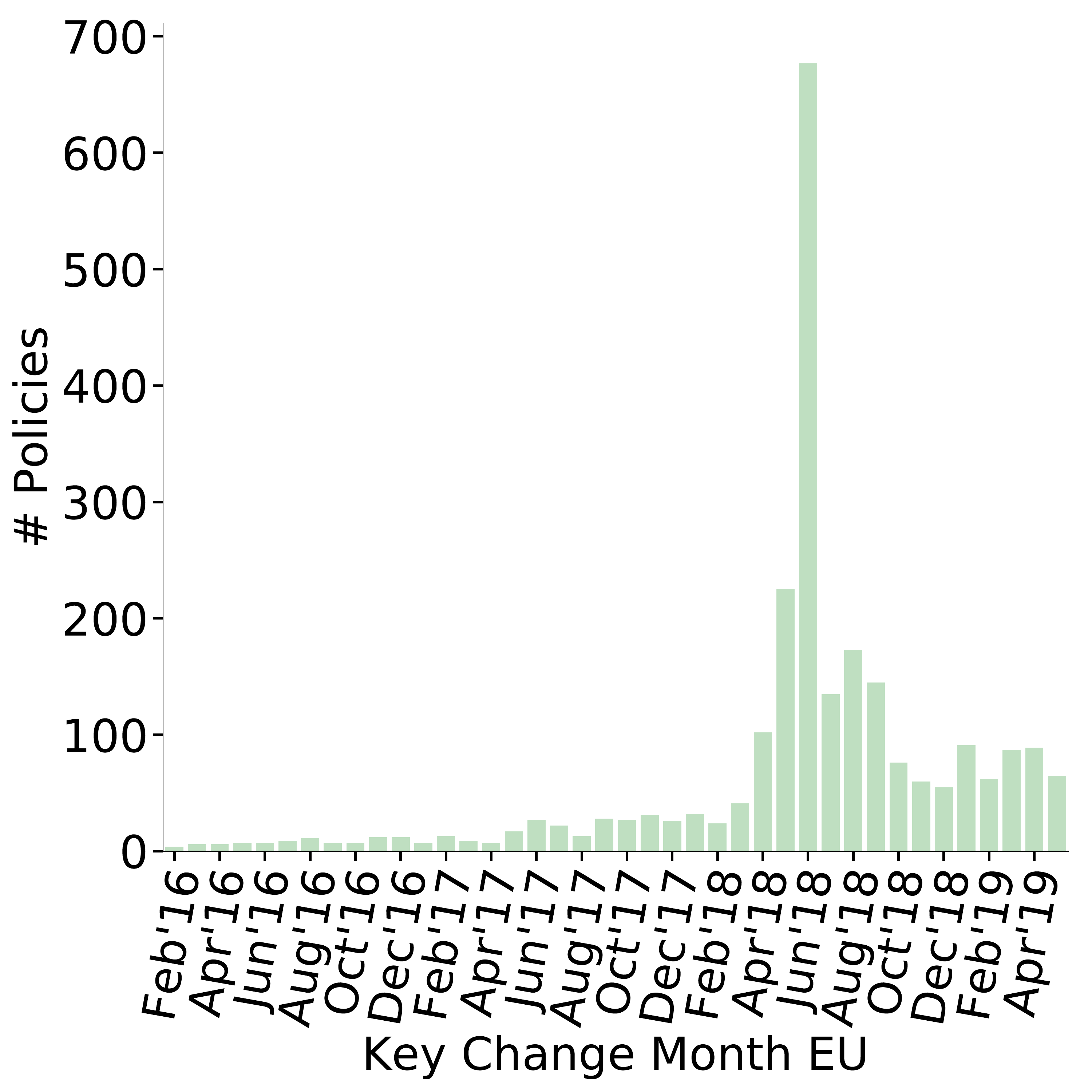}
         \caption{\textit{EU} dataset.}
         \label{fig:key_change_eu}
     \end{subfigure}
     \hfill
     \begin{subfigure}[b]{0.49\columnwidth}
         \includegraphics[width=\textwidth]{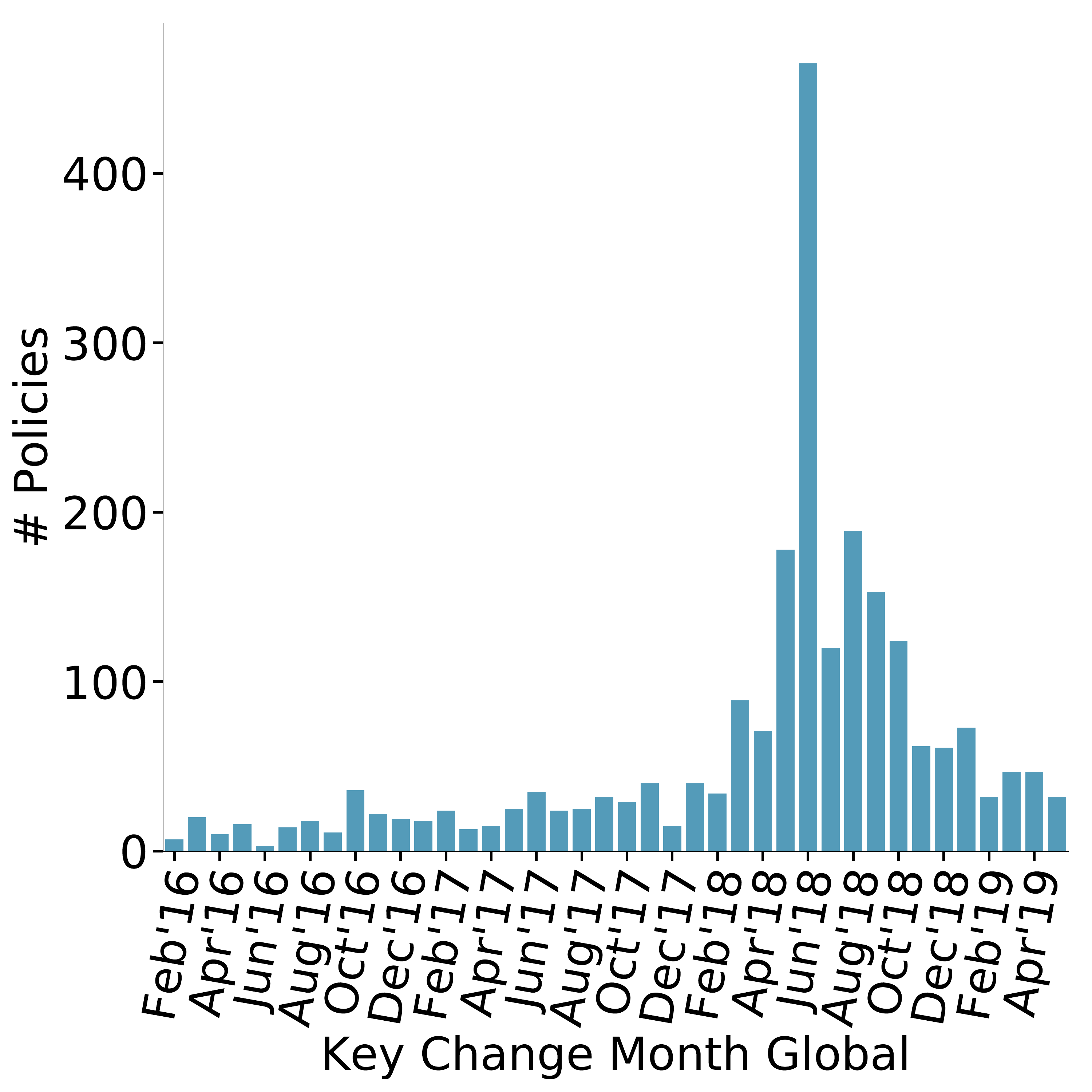}
         \caption{\textit{Global} dataset.}
         \label{fig:key_change_global}
     \end{subfigure}
        \caption{The distribution of key-change dates for the privacy policies in our corpus.}
        \label{fig:key_change}
\end{figure}

\subsubsection*{Analyzed Dataset}

Next, we used the text of the valid snapshots for each policy to examine its change over time. We quantified the per-month change using a fuzzy string matching library\footnote{\url{https://github.com/seatgeek/fuzzywuzzy}} that employs Levenshtein Distance to calculate the distance between two strings.
We defined changes in text between two snapshots to be \textit{significant} if the similarity ratio between the two text files was less than or equal to 95\%. For both datasets, a  portion of the policies exhibited no significant changes between 2016 and 2019. These policies have the same \pregdpr and \postgdpr versions. For the rest, we define, for each policy, the \textit{key-change date} as the closest month to the enforcement date of the GDPR that exhibited a significant change. The \pregdpr snapshot of the policy is the first stable version of the policy before the key-change date with a timestamp preceding May 2018. This strategy accommodates the transient stage around the enforcement date of the GDPR by considering a stable version of the policy. We used the most recent snapshot (taken after May 2018) of each policy as the \postgdpr version since as it captures any changes that may have occurred after the key-change date.
The distribution of key-change dates for the \textit{EU} and \textit{Global} sets can be seen in Fig.~\ref{fig:key_change_eu} and Fig.~\ref{fig:key_change_global}, respectively. The most frequent key-change date in both sets is June 2018, the month following the enforcement of the GDPR, as snapshots captured in this month contained the changes deployed in order to meet the deadline.

\begin{figure}[t]
    \centering
  \includegraphics[width=0.7\columnwidth]{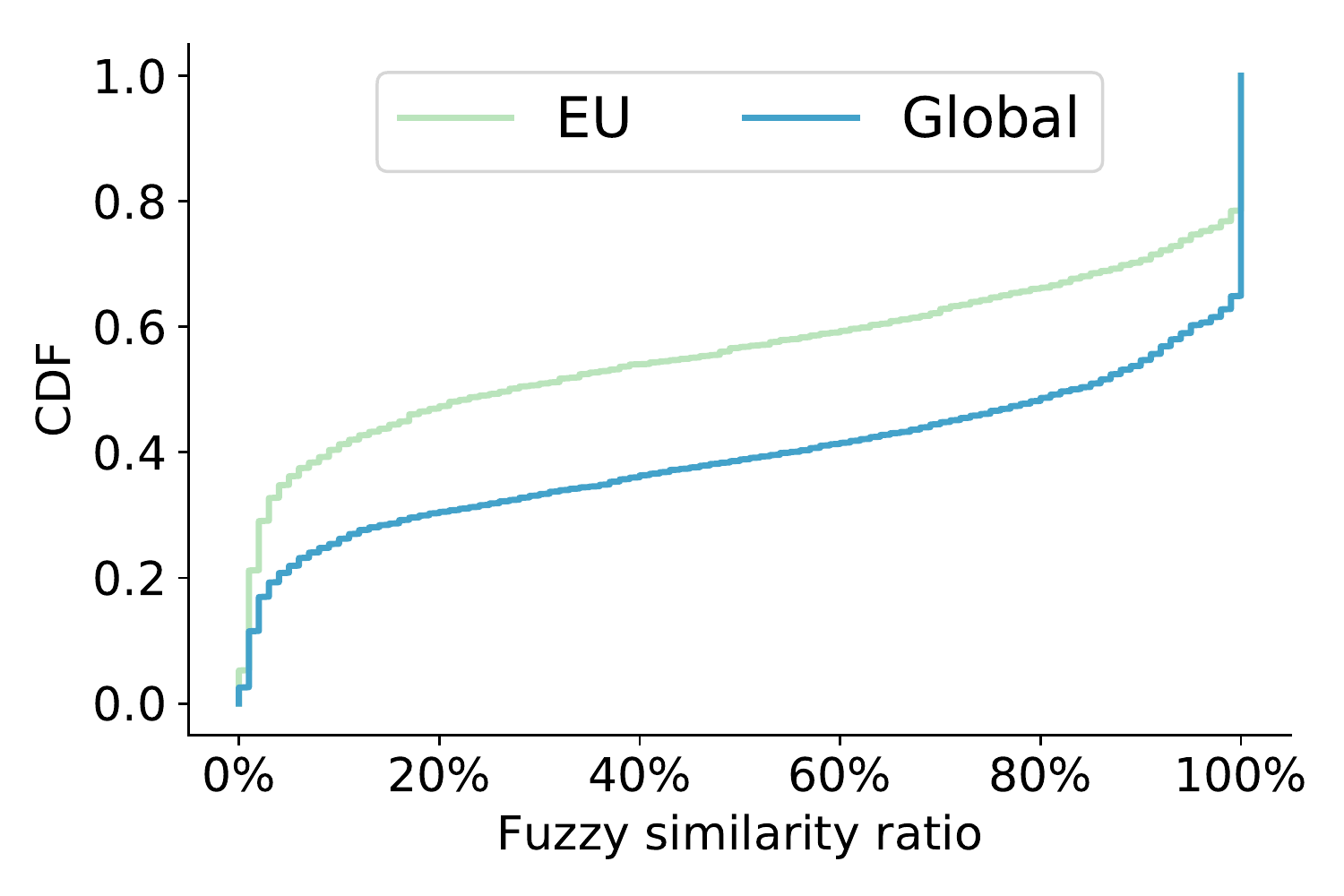}
    \caption{Cumulative distribution of the fuzzy similarity metrics for the \pregdpr and \postgdpr pairs of both datasets.}
    \label{fig:fuzzy_ratio_combined}
\end{figure}

We removed policies that lacked a valid \pregdpr or \postgdpr snapshot from further analysis (e.g., a policy was not indexed by the Wayback Machine before May 2018). This last pruning step resulted in the final \textit{EU} dataset of 3,084 pairs of snapshots and a \textit{Global} dataset of 3,592 pairs of snapshots. Between the two sets, 398 policy URLs are shared; these intersecting instances are globally popular sites (\textit{e.g. Microsoft and Dictionary.com}), along with instances of websites added to the \textit{Global} set as members of the ``Europe'' category (\textit{e.g. Times of Malta and Munster Rugby}).
Fig.~\ref{fig:fuzzy_ratio_combined} illustrates the cumulative distributions of fuzzy similarity between the \pregdpr and \postgdpr instances for the \textit{EU} and \textit{Global} sets. As evident from the figure, the \textit{EU} set demonstrates more signs of change during this period.

\subsection{Takeaways}
Our methodology highlights two takeaways regarding the initial impact of the GDPR on the privacy policy landscape. As expected, May 2018 was an inflection point; more than 45\% of the policies that we studied have changed between March and July 2018. Second, the GDPR appears to have had a higher impact inside the EU than outside. The text content of the policies corresponding to EU websites has changed at a higher rate than their global counterparts.

\section{Presentation Analysis}
\label{sec:presentation}
Our first step to understanding the evolution of the privacy policies is to test for changes in the presentation of the web pages through a user study. We followed a within-subjects study design, with two conditions: \pregdpr and \postgdpr. Our goal was to have each participant evaluate how presentable a screenshot of a privacy policy is.

\subsection{Hypothesis}

Our null hypothesis for the user evaluation of privacy policies is that \textit{there is no significant difference in users' perception of privacy policy appearance between the pairs of \pregdpr and \postgdpr policies in our sample.} We reject the null hypothesis for $p < .05$.

\subsection{Study Setup}

We recruited 470 participants (so that each snapshot receives at least ten different evaluations) from Amazon Mechanical Turk. We chose participants who had \mbox{$>95\%$} HIT approval rate and achieved masters status. We paid each respondent \$1.45 to fill the survey that lasted 8 minutes on average. Out of the respondents, 48\% were female, 46\% had a Bachelors degree, and 26\% did not have a degree. The average age of the respondents was 39 years. We did not ask for any personally identifiable information.

\subsubsection*{Study Material}

We chose a random set of 200 unique privacy policies from the \textit{EU} set and 200 unique privacy policies from the \textit{Global} set (we excluded policies common between the two sets). Counting the \pregdpr and \postgdpr versions, we obtained a total of 800 policies.
We followed the approach used in previous studies around websites aesthetics~\cite{reinecke2013predicting,lindgaard2006attention} to assess the presentability by using screenshots of the webpages instead of live versions. This approach avoids any bias from webpage loading times, internet speed or localized versions. 
We used the ``webkit2png'' tool to capture a full screenshot of each of the 800 privacy policies, which were all hosted by the Wayback Machine.
As these 800 screenshots included the full policy scrolled to the bottom, we cropped $612 \times 1028$ pixels from the text body of each screenshot to display for the respondents. Two of the authors manually inspected each of the images and corrected any screenshot that was incorrectly captured.

\subsubsection*{Survey Design}

We presented each respondent with a random set of 20 screenshots from the total set of 800 images.
The image order was randomized per participant to compensate for the effects of learning and fatigue. The respondents were not primed about the comparative nature of the study. 
We explicitly asked the respondents not to read the content of each screenshot, but instead to give their assessment over the images' look and feel. For each screenshot, the respondents indicated how much they agree/disagree with a set of three statements over a 5-point Likert scale (Strongly Disagree(SD), Disagree (D), Neither (N), Agree (A), Strongly Agree (SA)). A snapshot of the survey is available in Fig.~\ref{fig:survey_snap} of Appendix~\ref{sec:appndx_survey}. These statements, consistent with the usability measurement questions in~\cite{loiacono2002webqual}, were as follows:

\begin{enumerate}
    \item[] \text{\sone.} This policy has an attractive appearance.
    \item[] \text{\stwo.} This policy has a clean and simple presentation.
    \item[] \text{\sthree.} This policy creates a positive experience for me.
\end{enumerate}

Additionally, we placed two anchor questions that contain poorly formatted ``lorem ipsum'' text, using them to filter out respondents with low-quality answers. At the survey end, the respondents filled an optional demographics survey. After filtering out responses failing the quality-assurance questions, we analyzed only those images with at least five evaluations. The resulting image set was composed of 139 pairs of \textit{EU} policies and 136 pairs of \textit{Global} policies. On average, each screenshot received 7.8 evaluations.

\subsubsection*{Findings}

The distribution of scores can be seen in Tables \ref{tab:eu_user_study_agree} and \ref{tab:user_study_agree}. Because we want to compare users' overall perception of policy interfaces, we group scores into three categories: disagree (Strongly Disagree, Disagree), neutral (Neither Agree nor Disagree), and agree (Agree, Strongly Agree). Given three discrete outcomes, we apply the Chi-Squared Test between the \pregdpr and \postgdpr instances' distributions. For the \textit{EU} set, we reject the null hypothesis for $p=0.05$ for \sone ($0.039$) and \stwo ($0.040$), but fail to reject the null hypothesis for \sthree ($0.057$). On the other hand, for the \textit{Global} set, we fail to reject the null hypothesis for $p=0.05$ for all questions, with all three scores having $p > 0.4$. This result suggests that \textit{EU} policies improved their visual interfaces to be more attractive and simplified to further their support of the GDPR's call for clarity in policies.

 \begin{table}[t]
 \begin{center}
\vspace{0.5\baselineskip}
 \scriptsize
  
    \caption{The resulting \textit{EU} scores for the user study grouped by question  and study condition.}
    \label{tab:eu_user_study_agree}
    \begin{tabularx}{\columnwidth}{m{0.2in} m{0.65in} m{0.65in} m{0.65in} m{0.65in}}
      \textbf{Stmt} & \textbf{Condition} & Disagree (\%) & Neutral (\%) & Agree (\%) \\
      \midrule 
    \sone & \pregdpr \par \postgdpr & 32.5 \par 28.2 & 19.8 \par 20.7 & 47.7 \par 51.1 \\ \midrule
    \stwo & \pregdpr \par \postgdpr & 24.8 \par 20.9 & 16.1 \par 16.6 & 59.2 \par 62.5 \\ \midrule
    \sthree & \pregdpr \postgdpr & 29.5 \par 25.5 & 26.4 \par 28.0 & 44.0 \par 46.5 \\ 
    \end{tabularx}
    \end{center}
\end{table}

 \begin{table}[t]
\vspace{0.5\baselineskip}
 \scriptsize
  \begin{center}
    \caption{The resulting \textit{Global} scores for the user study grouped by question  and study condition.}
    \label{tab:user_study_agree}
    \begin{tabularx}{\columnwidth}{m{0.2in} m{0.65in} m{0.65in} m{0.65in} m{0.65in}}
      \textbf{Stmt} & \textbf{Condition} & Disagree (\%) & Neutral (\%) & Agree (\%) \\
      \midrule 
    \sone & \pregdpr \par \postgdpr & 35.0 \par 33.6 & 20.4 \par 20.0 & 44.6 \par 46.4 \\ \midrule
    \stwo & \pregdpr \par \postgdpr & 26.1 \par 25.5 & 18.4 \par 16.2 & 55.5 \par 58.3  \\ \midrule
    \sthree & \pregdpr \par \postgdpr & 32.5 \par 32.1 & 27.2 \par 29.2 & 40.2 \par 38.7 \\ 
    \end{tabularx}
    \end{center}
\end{table}

Another planned comparison was to investigate how the 
\pregdpr \textit{EU} policies compare to their \textit{Global} counterparts as well as how the \postgdpr sets compare. We group the scores into the same three categories as before, and once again apply the Chi-Squared Test.  
We fail to reject the null hypothesis for $p=.05$ for all three questions when comparing the \pregdpr sets; however, we reject the null-hypothesis for \sone ($p=0.007$), \stwo ($p=0.014$), and \sthree ($p=9.06e-5$) when comparing the \postgdpr images of the two sets.
Looking at the \postgdpr distributions in Table.~\ref{tab:eu_user_study_agree} and Table.~\ref{tab:user_study_agree}, the observable difference between the two sets of scores suggests that the \textit{EU} policies created a notably more positive experience for users compared to the \textit{Global} policies.

\subsection{Takeaways}
    The major takeaway from this study is that the GDPR was a driver for enhancing the appearance and the presentation of the top privacy policies in the EU countries. We were not able to conclusively observe such an improvement for the \textit{Global} set of websites. Compared to the  \textit{Global} set, the \textit{EU} websites have improved the appearance, presentation, and experience of their policies. 
  
\section{Text-Feature Analysis}
\label{sec:readability}

We complement our initial study of the visual features' evolution in the privacy policies with an assessment of syntactic textual changes. These features allow us to get a high-level understanding of how the policy's structure evolved before we go deeper into semantic features in the later sections.

\subsection{Hypothesis}

We applied five common text metrics to the documents that describe the length and sentence structure of the text.  For each test we conduct: we evaluate the null-hypothesis that \textit{there does not exist a significant change in the tested text-feature between the \pregdpr 
and 
\postgdpr
instances of a policy}. Since we run five tests on the same policy samples, we apply the Bonferroni correction~\cite{bonferroni}. Below, the null hypothesis is rejected when the $p$ value of the test is less than $\frac{0.05}{5}$.

\subsection{Text Metrics}

We consider the following metrics:

\begin{enumerate}
\item
     \textbf{Syllables Count:} gives the total number of syllables available in the text.
\item
     \textbf{Word Count:} gives the number of words available in the text.
\item
     \textbf{Sentence Count:} gives the number of sentences present in a text.
\item
     \textbf{Word per Sentence:} gives the average number of words per sentence in the text.
\item
 \textbf{Passive Voice Index:} gives the percentage of sentences that contain passive verb forms.
To compute this score, we tokenize the text into sentences and perform dependency parsing on each sentence using the Spacy library\footnote{https://spacy.io/}. We consider a sentence to contain a passive voice if it follows the pattern of: \mbox{\textsl{nsubjpass}} (that is Nominal subject (passive)), followed by \mbox{\textsl{aux}} (Auxiliary), and then followed by \mbox{\textsl{auxpass}} (Auxiliary (passive)). This pattern would match sentences similar to \textit{``Data is collected.''} 
\end{enumerate}

\subsection{Findings}

\begin{table}[t]
 \scriptsize
  \renewcommand{\arraystretch}{1.1}
   \begin{center}
    \caption{EU Text Metrics results for \pregdpr and \postgdpr instances.}
    \label{tab:eu_readability_stats}
    \begin{tabularx}{\columnwidth}{l l l l}
      \textbf{Metric} & \textbf{\pregdpr} $\mu$ ($\sigma$) & \textbf{\postgdpr} $\mu$ ($\sigma$) & \textbf{$p$} \\
      \midrule 
        $\#$Syllables & 3534.18 (4580.42) & 4670.45 (5291.80)& 9.92e-143\\
        $\#$Words & 1936.15 (2010.29) & 2621.38 (2564.93) & 3.05e-148 \\
        $\#$Sentences & 53.49 (53.35) & 71.36 (69.62) & 1.02e-100  \\
        $\#$Words/Sent.  & 53.23 (121.10) & 50.47 (144.46)& 2.84e-06\\
        $\#$Passive & 10.75 (7.35) & 10.90 (6.89) & 1.85e-01 \\
    \end{tabularx}
   \end{center}
\end{table}

\begin{table}[t]
 \scriptsize
  \renewcommand{\arraystretch}{1.1}
   \begin{center}
    \caption{Global Text Metrics results for \pregdpr and \postgdpr instances.}
    \label{tab:readability_stats}
    \begin{tabularx}{\columnwidth}{l l l l}
      \textbf{Metric} & \textbf{\pregdpr} $\mu$ ($\sigma$) & \textbf{\postgdpr} $\mu$ ($\sigma$) & \textbf{$p$} \\
      \midrule 
        $\#$Syllables & 2977.43 (2941.52) & 4108.18 (6606.61)& 6.36e-136\\
        $\#$Words & 1709.45 (1609.23) & 2140.44 (2167.40) & 6.16e-141 \\
        $\#$Sentences & 48.04 (42.02) & 58.61 (54.42) & 1.52e-86  \\
        $\#$Words/Sent.  & 42.47 (47.61) & 43.76 (54.51)& 4.14e-04\\
        $\#$Passive & 11.45 (7.21) & 11.52 (6.99) & 4.29e-01 \\
    \end{tabularx}
   \end{center}
\end{table}

We compute the value of each text metric for the \pregdpr and \postgdpr versions of each policy. Given the high variability in the text content of policies, we avoid using statistics requiring normal-distributions. Instead, we use the Wilcoxon signed rank test to compare the \pregdpr and \postgdpr pairs of text for each metric, correcting for multiple comparisons. Table~\ref{tab:eu_readability_stats} and Table~\ref{tab:readability_stats} describe the key statistics of the study for the \textit{EU} and \textit{Global} sets, respectively.

For the \textit{EU} set, we reject the null hypothesis for the changes in the number of syllables (+50\%), the number of words (+35\%), and the number of sentences (+33\%). The magnitude of these changes indicates that policies in this set are becoming much longer. These results are consistent with the expected effects of adopting the GDPR. \textit{EU} policies must cover more material than before to bring transparency to data subjects.

For the \textit{Global} set, there has been a statistically significant increase in the number of syllables (+38\%), the number of words (+25\%), and the number of sentences (+21\%). 
This increase in the size of policies outside the EU suggests that the GDPR has significantly impacted the global landscape of privacy policies.
Still, the magnitude of this increase is noticeably smaller compared to the \textit{EU} set. 
Finally, we fail to reject the null hypothesis for the passive voice index as with the \textit{EU} set. 

\subsection{Takeaways}
Consistent with recent findings~\cite{nytimes_2019, techcrunch_2019}, we find privacy policies to have increased in length considerably, with \textit{EU} policies being longer than \textit{Global} policies. We, however, find that this increase did not accompany sentence structure improvements. The passive voice index did not exhibit any change in the \postgdpr versions for both the datasets. Further, while there are statistically significant changes in the number of words per sentence, these changes are too small (-5\% for \textit{EU} and +3\% for \textit{Global}) to conclude a change in the sentence structure of the policies. On average, a privacy policy sentence spans more than 43 words for both datasets.

\section{Automated Requirement Analysis}
\label{sec:reqs}
While the text-feature analysis captures whether the privacy policies have syntactically changed, the metrics are domain-agnostic. Hence, we follow a different methodology to delve deeper and assess the coverage, the compliance, and the specificity angles in the following sections.

 \subsection{Methodology Overview}
\begin{figure}[t]
    \centering
  \includegraphics[width=1\columnwidth]{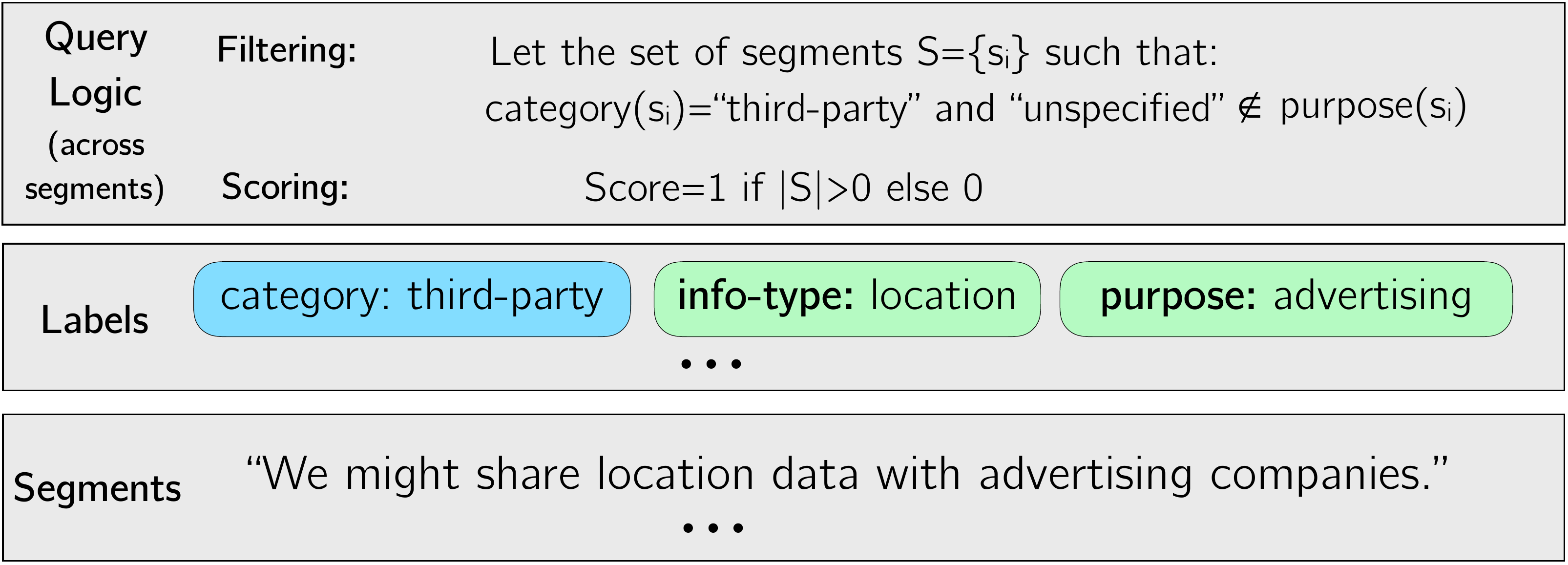}
    \caption{Structured querying with two-levels of abstraction on top of text}
    \label{fig:structured}
\end{figure}

Our methodology starts by defining a set of goals or requirements. 
These goals are high-level and are independent of the implementation methodology. 
Then, we code these goals by extending a technique called \textit{structured querying} over privacy policies, introduced by Harkous et al.~\cite{harkous2018polisis}. 
This technique builds two levels of abstraction on top of the raw privacy policy as demonstrated in Fig.~\ref{fig:structured}. 

On the first level, a privacy policy is converted from a set of text segments written in natural language to a set of automatic labels that describe the embedded privacy practices. For instance, one label can indicate that the segment discusses ``third-party'' sharing and another can indicate that the sharing purpose is ``advertising''. These labels are assigned by a set of machine-learning text classifiers trained on human annotations.

On the second level, a first-order logic query is constructed to reason about the set of labels across segments. For example, if one were to query the level of specificity of a policy describing the purpose of data collection, they would list the segments that have a purpose label. Then, they would count how many of these segments do not have the purpose label equal to ``unspecified.'' We further formalize the querying logic in this work by splitting the querying into two steps. The first step is \textbf{filtering}, where the first-order logic query is used to filter the total set of segments into a subset of relevant segments. The second step is \textbf{scoring}, where a scoring function is computed based on the relevant segments.
We provide an example in Fig.~\ref{fig:structured}, where the filtering step is used to decide on the subset of segments discussing third-party sharing with specified purposes and the scoring step assigns a score of 1 if this subset is not empty.
We follow the spirit of this \textbf{filtering-scoring} approach for our in-depth analysis in the following sections.

This structured querying technique offers an advantage over approaches based on heuristics and keyword analysis (e.g.,~\cite{anton2002analyzing, degeling2018we}); it allows us to better cover text with varying wordings but similar semantics. Further, this technique avoids the shortcomings of the other approaches that directly use machine learning to quantify the goals (e.g.,~\cite{contissa2018claudette, lippi2018claudette,lebanoff2018automatic}); it is more flexible for adapting the goals (i.e., queries) as needed, without having to create new labeling data for each new goal.

In this work, we are the first to conduct a comprehensive analysis of privacy-related goals using structured querying. Our main contributions lie in the goals' definition, the translation of these goals into queries, the volume of goals we measure, and the comparative nature of this study.

\subsection{Polisis}

\begin{figure*}[t]
    \centering
  \includegraphics[width=1\textwidth]{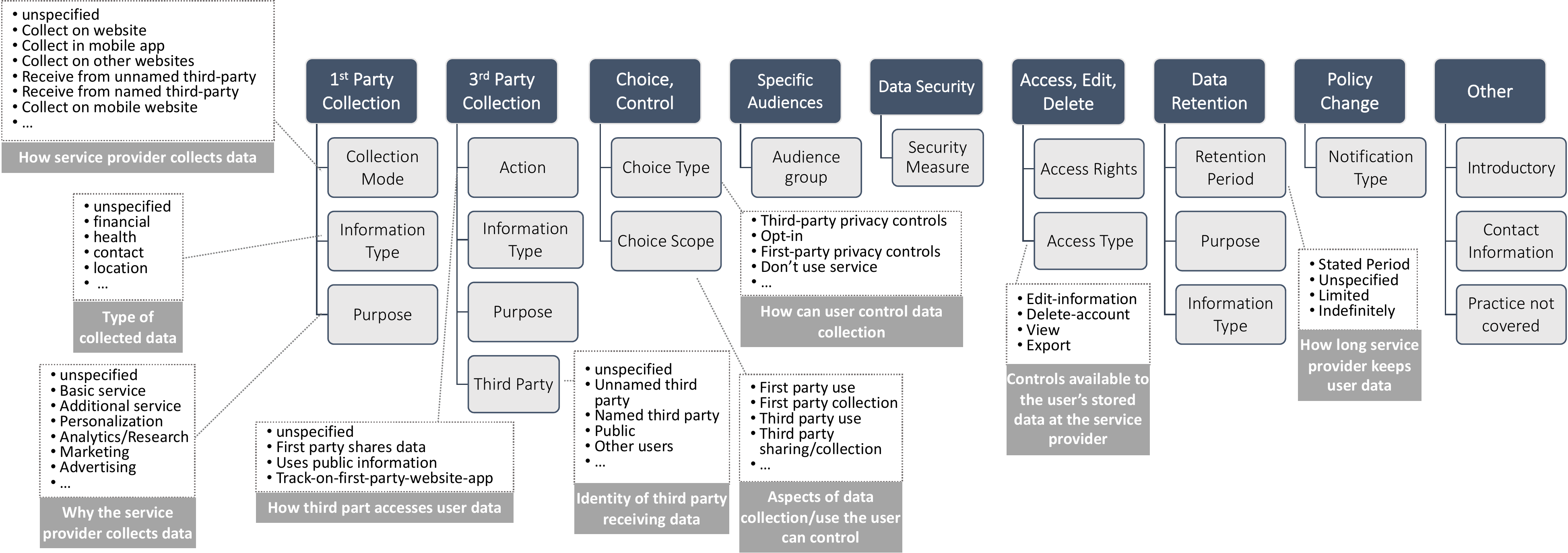}
    \caption{ The privacy taxonomy of Wilson {\em et al.}~\cite{Wilsonacl16}. The top level of the hierarchy (darkly shaded blocks) defines high-level privacy categories. The lower level defines a set of privacy attributes (light shaded blocks), each assuming a set of values. We show examples of values for some of the attributes. The taxonomy has more attributes that we do not show for space considerations.}
    \label{fig:opp}
\end{figure*}

We utilize the Polisis system described by Harkous et al~\cite{harkous2018polisis} to generate the automated labels described within structured querying. 
Polisis pre-processes a privacy policy and breaks it into a set of smaller segments (one example of such a segment is in Fig.~\ref{fig:structured}). A segment is a set of consecutive sentences of a privacy policy that are semantically coherent. 
Polisis passes each segment through a set of classifiers to assign automatic labels describing the embedded privacy practices. 
These classifiers have been trained on the OPP-115 dataset created by Wilson et al.~\cite{Wilsonacl16}. The dataset consists of 115 privacy policies (267K words) with manual annotations for 23K fine-grained data practices. The privacy-policy taxonomy used for the annotation task is depicted in (Fig.~\ref{fig:opp}). 

Polisis labels each segment with high-level privacy categories (blue labels in Fig.~\ref{fig:structured}) as well as values for lower-level privacy attributes (green labels in Fig.~\ref{fig:structured}). 
In particular, Polisis assigns a segment $s$, of a privacy policy, a set: $\category(s)$. This set is a subset of the nine high-level privacy categories which are dark shaded in Fig.~\ref{fig:opp}. Also, Polisis labels each segment with a set of values, corresponding to 20 lower-level privacy attributes (light-shaded in Fig.~\ref{fig:opp}). The values corresponding to each attribute are shown as tables in Fig.~\ref{fig:opp}. For example, the attribute ``purpose'' indicates the purposes of data processing and is represented by the set $\purpose(s)$.

If $\category(s) = \{ \first \}$ and $\purpose(s)=\{
\textsl{basic-feature},\; \textsl{personalization},\; \textsl{marketing}\}$, we conclude that the segment $s$ describes multiple purposes for first party data collection, which are to provide basic features, personalize the service, and use data for marketing. In addition to the labels, Polisis returns a probability measure associated with each label. Elements of the sets mentioned above are the ones classified with a probability larger than 0.5. 

Following guidelines from the developers of Polisis, we retrained its classifiers so that they can execute locally. We refer the reader to the work of Harkous et al.,~\cite{harkous2018polisis} for a breakdown of the accuracy of the high-level and each of the low-level classifiers. 
Also, a detailed description of all the taxonomy attributes and their values is present within the OPP-115 dataset (\url{https://usableprivacy.org/data}).

\begin{table}[t]
 \scriptsize
  \begin{center}
    \caption{Description of the relevant high-level privacy categories from Wilson et al.~\cite{Wilsonacl16}.}
    \label{tab:table1}
    \begin{tabularx}{\columnwidth}{m{2.5cm} m{5.5cm}}
      \textbf{Privacy Category} & \textbf{Description}\\
      \midrule
      First Party Collection /Use & Service provider's collection and use of user data.\\
\hline
Third Party Sharing /Collection  &  Sharing and collection of user data with third parties (e.g., advertisers)\\
\hline
User Choice/Control & Options for choices and control
users have for their collected data \\
\hline
International \& Specific Audiences & Practices
related to a specific group of users
(e.g., children, Europeans).
       \\
       \hline
       Data Security & The protection mechanisms for user's data. \\
\hline
User Access, Edit, \& Deletion & Options for users to
access, edit or delete their stored data. \\
\hline
Data Retention & The period and purpose of storing user's data. \\
\hline
Policy Change & Communicating changes to the privacy policy to the users.\\
\hline
Privacy Contact Info & The contact information for privacy related matters.\\
    \end{tabularx}
  \end{center}
\end{table}

\section{Coverage Analysis}
\label{sec:cat_coverage}

As the first step in the context-based analysis of privacy policies, we study the policies' coverage of the high-level privacy categories described in Table~\ref{tab:table1}. This analysis highlights the difference in category coverage between \pregdpr and \postgdpr policies. A policy covers a specific category if it contains at least one segment with this category as a label, according to Polisis.

\begin{figure}[bh!]
    \centering
  \includegraphics[width=0.9\columnwidth]{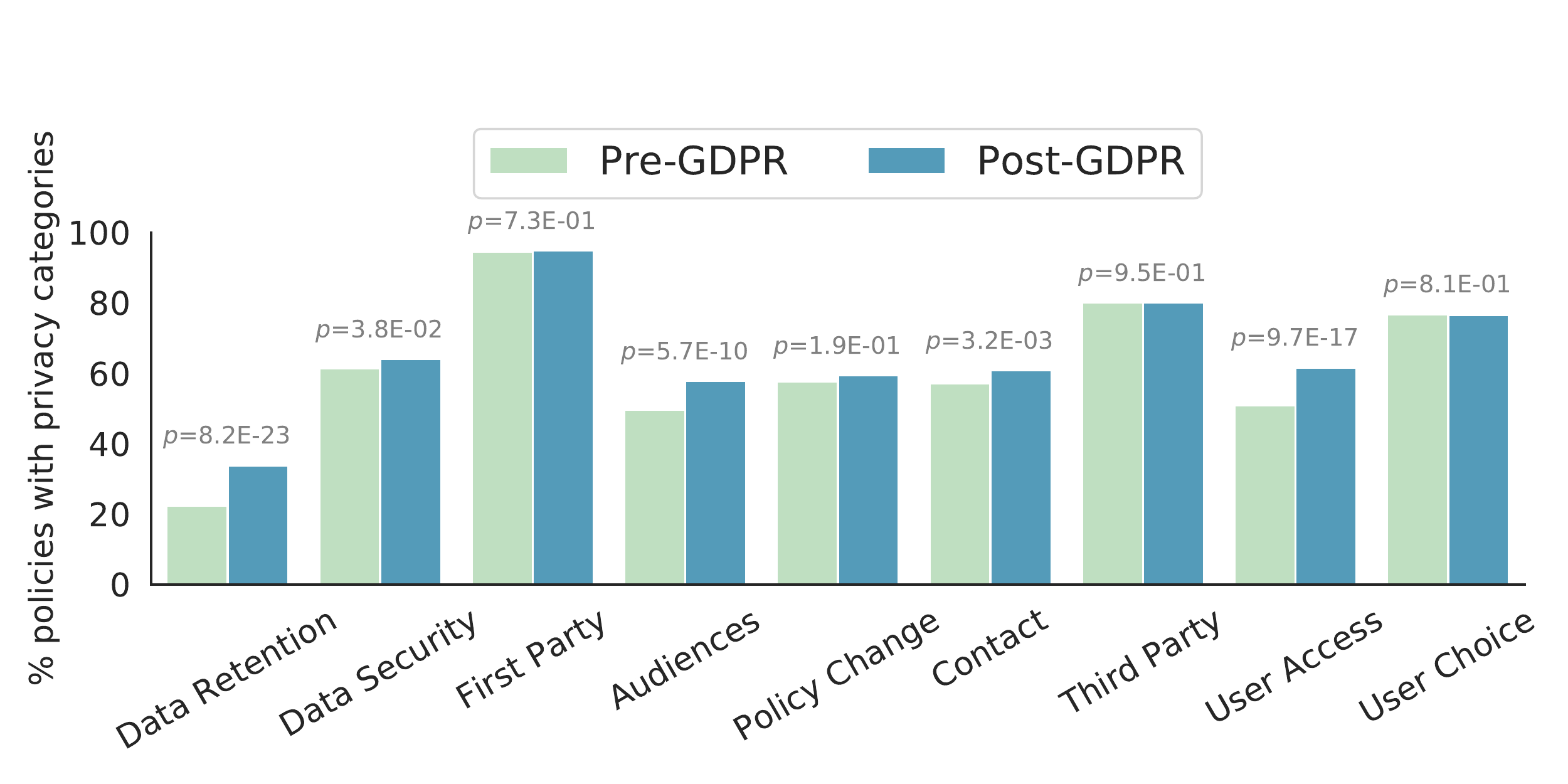}
    \caption{Category Coverage for \textit{EU} policies' \pregdpr and \postgdpr instances; $p$ values are derived from applying the Chi-Square test. }
    \label{fig:eu_cat_coverage}
    \vspace{-0.5\baselineskip}
\end{figure}

\begin{figure}[th!]
    \centering
  \includegraphics[width=0.9\columnwidth]{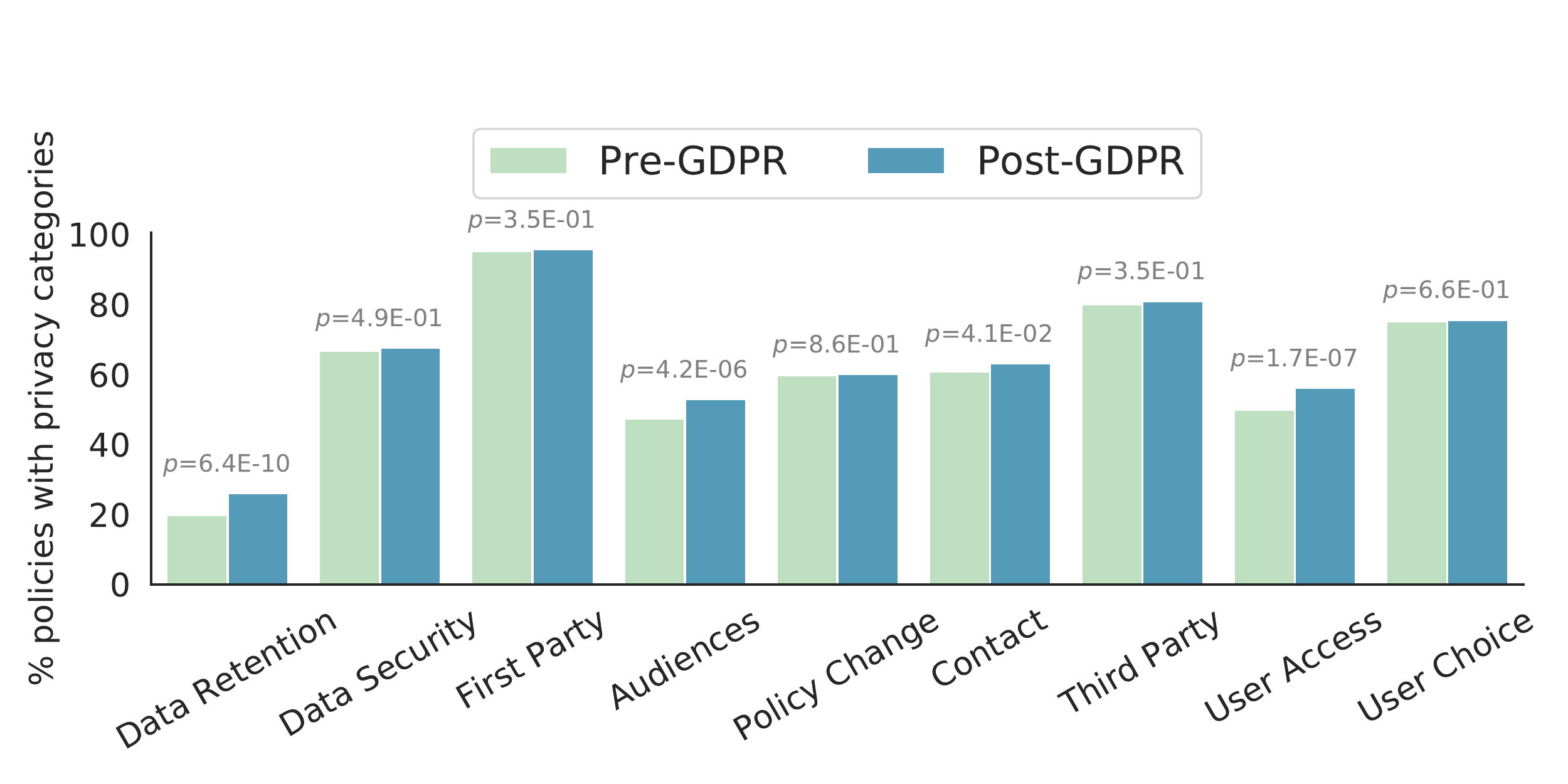}
    \caption{Category Coverage for \textit{Global} policies' 
    \pregdpr and \postgdpr instances; $p$ values are derived from applying the Chi-Square test. }
    \label{fig:cat_coverage}
\end{figure}

Fig.~\ref{fig:eu_cat_coverage} and Fig.~\ref{fig:cat_coverage} display the fraction of \textit{EU} and \textit{Global} policies with coverage scores of 1 for each of the high-level privacy categories \pregdpr and \postgdpr. 
For this analysis, we consider the hypothesis that \textit{there is no significant change in the category coverage of privacy policies between \pregdpr and \postgdpr instances}. We use the Chi-Squared test to evaluate this hypothesis. However, since we are doing multiple tests on the same labeled data set, we apply Bonferroni correction \cite{bonferroni} and reject the null hypothesis only when the $p$ value is less than $ \frac{.05}{9} $. The $p$ values for each category are shown in Fig.~\ref{fig:eu_cat_coverage} and Fig.~\ref{fig:cat_coverage} for the \textit{EU} and \textit{Global} sets, respectively. 
    
We observe that categories already exhibiting high coverage did not have a major change between the \pregdpr and the \postgdpr sets for both the \textit{EU} and the \textit{Global} cases. On the other hand, categories that were under-represented in the \pregdpr set showed significant change. For the \textit{EU} set, we observe statistically significant improvement in four of the nine categories  (`Data Retention', `International \& Specific Audiences', `Privacy Contact Information', and `User Access, Edit \& Deletion'). For the \textit{Global} set, we observe statistically significant improvement in three of the nine categories (Data Retention, International \& Specific Audiences, and User Access, Edit \& Deletion).

\subsection{Comparison to Manual Annotations} \label{sec:manual_analysis}
As mentioned in Sec.~\ref{sec:reqs}, this context-based analysis uses Polisis~\cite{harkous2018polisis} as the core querying engine. A question that comes to mind is: \textit{How well does Polisis work with these queries?} To better understand how Polisis behaves at the query level, we decided to compare its results with Subject Matter Experts (SMEs).
We leverage the raw annotations of the OPP-115 dataset~\cite{Wilsonacl16} in which three SMEs have annotated each policy. 

After consulting with the authors of Polisis, we selected a set of 15 policies outside the training and validation sets of Polisis. We then generated the automatic labels for these policies.
Therefore, we ended up with four annotations of the same policy; three by SMEs and one by Polisis. 
We then passed these annotations through the querying engine to get the query results.
Next, we computed the level of disagreement in these results among SMEs themselves and between Polisis and SMEs. The disagreement rate per policy is quantified by the ratio of (the number of queries with different scores between the two groups) to (the total number of queries). We then average this rate across policies\footnote{We use this metric vs. Cohen's Kappa or Krippendorff's alpha as the latter does not apply for when the three annotators are not the same across policies, which was the case with the OPP-115 dataset.}.

We find that the disagreement rate for Polisis-SMEs was 0.10, which is only slightly worse than the SME-SME disagreement rate of 0.07. This observation indicates that we can rely on Polisis' annotations as a reasonable proxy to what human annotators produce. Note that the disagreement rate is not equivalent to the error rate as the latter assumes the existence of ground truth. 

\subsection{Takeaways}
It is evident from Fig. \ref{fig:eu_cat_coverage} and Fig. \ref{fig:cat_coverage} that the GDPR has had a positive effect on the coverage of categories.
 Traditionally, privacy policies covered clauses mostly related to first party collection and third party sharing. With the introduction of the GDPR, it is clear that there is a trend of covering additional categories of particular importance to the GDPR requirements, including data retention periods (Article 13(2.a)), notices to special audiences, safeguarding the user data, and providing the users with the options to access and rectify their information (c.f. Sec.~\ref{sec:background}).
Interestingly, the improvement in coverage of \textit{EU}  policies for these categories is higher than that of \textit{Global} policies. Finally, our manual analysis with subject matter experts shows that automatically extracting these results did not hinder their quality.

\section{Compliance Analysis}
\label{sec:compliance}
\begin{table*}[t]
 \scriptsize
  \begin{center}
    \caption{The list of the queries derived from ICO's GDPR checklists. ICO-Q1 -- ICO-Q7 are from the ``Right to be Informed'' checklist. ICO-Q8 is from the ``Right of Access'' checklist.
  { \scriptsize $S_{\textit{actions}}$=  \{collect-from-user-on-other-websites,
receive-from-other-parts-of-company-affiliates,
receive-from-other-service-third-party-named,
receive-from-other-service-third-party-unnamed,
track-user-on-other-websites}
\}}
    \label{tab:ico-qs}
    \begin{tabularx}{\textwidth}{m{2.0in} m{0.5in} m{3.0in} m{1.4in}}
      \textbf{ICO Checklist Item} &\textbf{GDPR Ref.} & \textbf{Filtering Logic} & \textbf{Scoring Func.}\\
      \midrule
      \textbf{ICO-Q1}: ``The purposes of processing user data.''
      &
        13(1.c)
      &
    Consider the set $S ={\{s_i\}}$ such that \par
    \makebox[3.0cm]{\category($s_i$)}=  \{\firsts\}
    \par
    {\textit{purpose}($s_i$)}$\neq \phi$ and 
     {\unspec} $\notin$ \textit{purpose}($s_i$)  
      &
      Score= $1$ \mbox{ if } \par $|S|>0$ \mbox{ else } $0$
      \\
      \midrule
      \textbf{ICO-Q2}: ``The categories of obtained personal data (if personal data is not obtained from the individual it relates to).''
      &
        14(1.d)
      &
    Consider the set $S ={\{s_i\}}$ such that \par
    \makebox[3.0cm]{\category($s_i$)}=  \{\firsts\}
    \par
    \makebox[3.0cm]{\textit{action-first-party}($s_i$)} $\subset S_{\textit{actions}}$ 
     \par
      \makebox[3.0cm]{\unspec} $\notin$ \textit{\pinfo}($s_i$)  
      &
      Score= $1$ \mbox{ if } \par $|S|>0$ \mbox{ else } $0$
      \\
      \midrule
        \textbf{ICO-Q3}: ``The recipients of the user's personal data.''
      &
        14(1.e)
      &
    Consider the set $S ={\{s_i\}}$ such that \par
    \makebox[3.0cm]{\category($s_i$)}=  \{\thirds\}
     \par
      \makebox[3.0cm]{\unspec} $\notin$ \textit{third-party-entity}($s_i$)  
      &
      Score= $1$ \mbox{ if } \par $|S|>0$ \mbox{ else } $0$
      \\
      \midrule
      \textbf{ICO-Q4}: ``The right for the user to withdraw consent from data processing.''
      &
        17(1.b)
      &
    Consider the set $S ={\{s_i\}}$ such that \par
    \makebox[3.0cm]{\category($s_i$)} $\in$  \{\firsts, \textit{user-choice-control}\}
     
    \makebox[3.0cm]{\textit{choice-type}($s_i$) }\hspace{-0.6cm} = {\scriptsize\{op-out-link, op-out-via-contacting-company\}}
      \par
    \makebox[3.0cm]{\textit{choice-scope}($s_i$) } = \{first-party-use\}
      &
      Score= $1$ \mbox{ if } \par $|S|>0$ \mbox{ else } $0$
      \\
      \midrule
      \textbf{ICO-Q5}: ``The source of the personal data (if the personal data is not obtained from the individual it relates to).''
      &
        15(1.g)
      &
    Consider the set $S ={\{s_i\}}$ such that \par
    \makebox[2.9cm]{\category($s_i$)}=  \{\firsts\}
    \par
    \makebox[2.9cm]{\textit{action-first-party}($s_i$)} $\subset$ 
    $S_{\textit{actions}}$
      &
      Score= $1$ \mbox{ if } \par $|S|>0$ \mbox{ else } $0$
      \\
      \midrule
      \textbf{ICO-Q6}: ``If we plan to use personal data for a new purpose, we update our privacy information and communicate the changes to individuals before starting any new processing.''
      &
        13(3)
      &
    Consider the set $S ={\{s_i\}}$ such that \par
    \makebox[3.0cm]{\category($s_i$)} =  \{\textit{policy-change}\}
    \par
    \makebox[3cm]{\textit{type-of-policy-change}($s_i$)} = \{privacy-relevant-change \}
        \par
    \makebox[3.0cm]{\unspec} $\notin$ {\textit{how-notified}($s_i$)}
      &
     Score= $1$ \mbox{ if } \par $|S|>0$ \mbox{ else } $0$
      \\
      \midrule
        \textbf{ICO-Q7}: ``Individuals have the right to access their personal data.'' 
      &
        15(1)
      &
    Consider the set $S ={\{s_i\}}$ such that \par
    \makebox[3.0cm]{\category($s_i$)} =  \{\textit{user-access-edit-deletion}\}
    \par
    \makebox[3.0cm]{\textit{access-type}($s_i$)} $\in$ \{view, export, edit-information\}
      &
     Score= $1$ \mbox{ if } \par $|S|>0$ \mbox{ else } $0$
        \\
    
    \end{tabularx}
  \end{center}
\end{table*}

Next, we study the content of the policies in light of the compliance requirements introduced by the GDPR. We rely on the UK's Information Commissioner's officer's (ICO) guide to the GDPR\footnote{\url{https://ico.org.uk/for-organisations/guide-to-the-general-data-protection-regulation-gdpr/}}, which contains a set of guidelines for organizations to meet the provisions set in the GDPR. 
In the form of a checklist for organizations to inform users of their rights, the ICO guide provides an official and structured interpretation of the GDPR. It obviates the need for our customized interpretation of the law. 
We translate these requirements via the filtering-scoring approach of Sec.~\ref{sec:reqs} in a way that allows us to compare the privacy practices of the service providers before and after the introduction of the GDPR. Table~\ref{tab:ico-qs} shows the ICO checklist items, their descriptions, and their corresponding filtering and scoring logic. 

Since the taxonomy employed in Polisis precedes the GDPR, some of the items in the ICO's checklists are incompatible with the framework, i.e., they cover newer concepts and are not quantifiable in the current framework. We considered only those items from the ICO framework that are compatible with Polisis taxonomy. We manually compared these covered items to the GDPR text to find them representative of articles 12-20. We consider the compliance evidence for all the checklist items as a binary metric denoted by the existence of a segment satisfying the associated clause.

To assess the change of compliance for each policy, according to these seven requirements, we compare the scores of \pregdpr and \postgdpr versions by breaking down the change for each requirement into four cases:

\begin{itemize}
    \item \textbf{Requirement Worsened:} A policy used to cover the requirement in the \pregdpr version, but does not cover it in the \postgdpr version; i.e., score drops from $1$ to $0$.
    \item \textbf{Requirement Still Missing:} A policy did not cover the requirement in either the \pregdpr or \postgdpr snapshots; i.e., the score is 0 for both versions.
    \item \textbf{Requirement Still Covered:} A policy continues to cover the requirement in both the versions; i.e., the score is 1 for both versions.
    \item \textbf{Requirement Improved:} A policy did not cover the requirement in the \pregdpr version, but does cover it in the \postgdpr version; i.e., the score rises from $0$ to $1$.
\end{itemize}

\begin{figure}[t!]
    \centering
  \includegraphics[width=1\columnwidth]{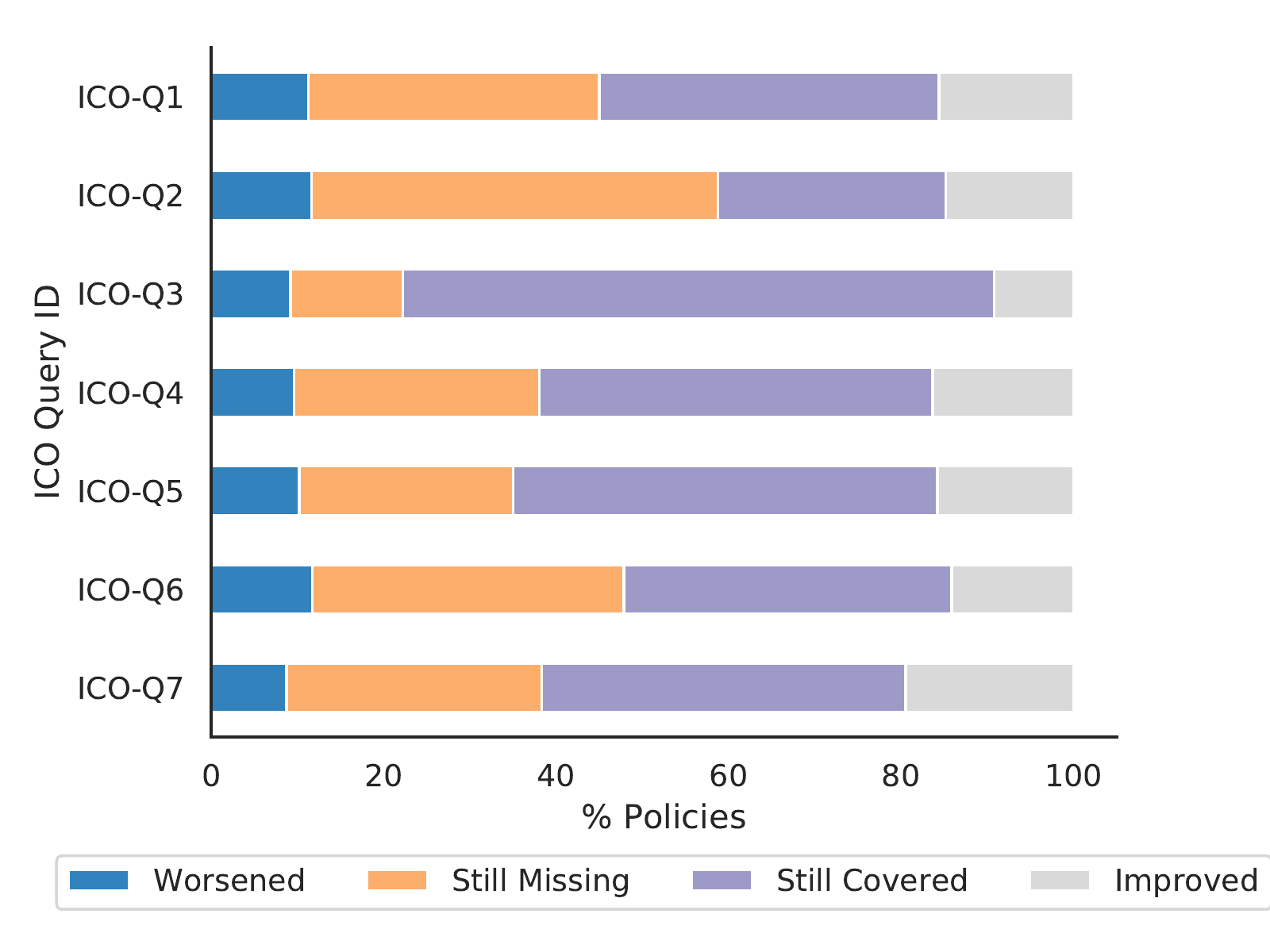}
    \caption{The comparison of ICO scores of \textit{EU} policies' \pregdpr and \postgdpr instances. The queries for the ICO checklist can be found in Table~\ref{tab:ico-qs}.}
    \label{fig:eu-ico-total}
\end{figure}

\begin{figure}[t!]
    \centering
  \includegraphics[width=1\columnwidth]{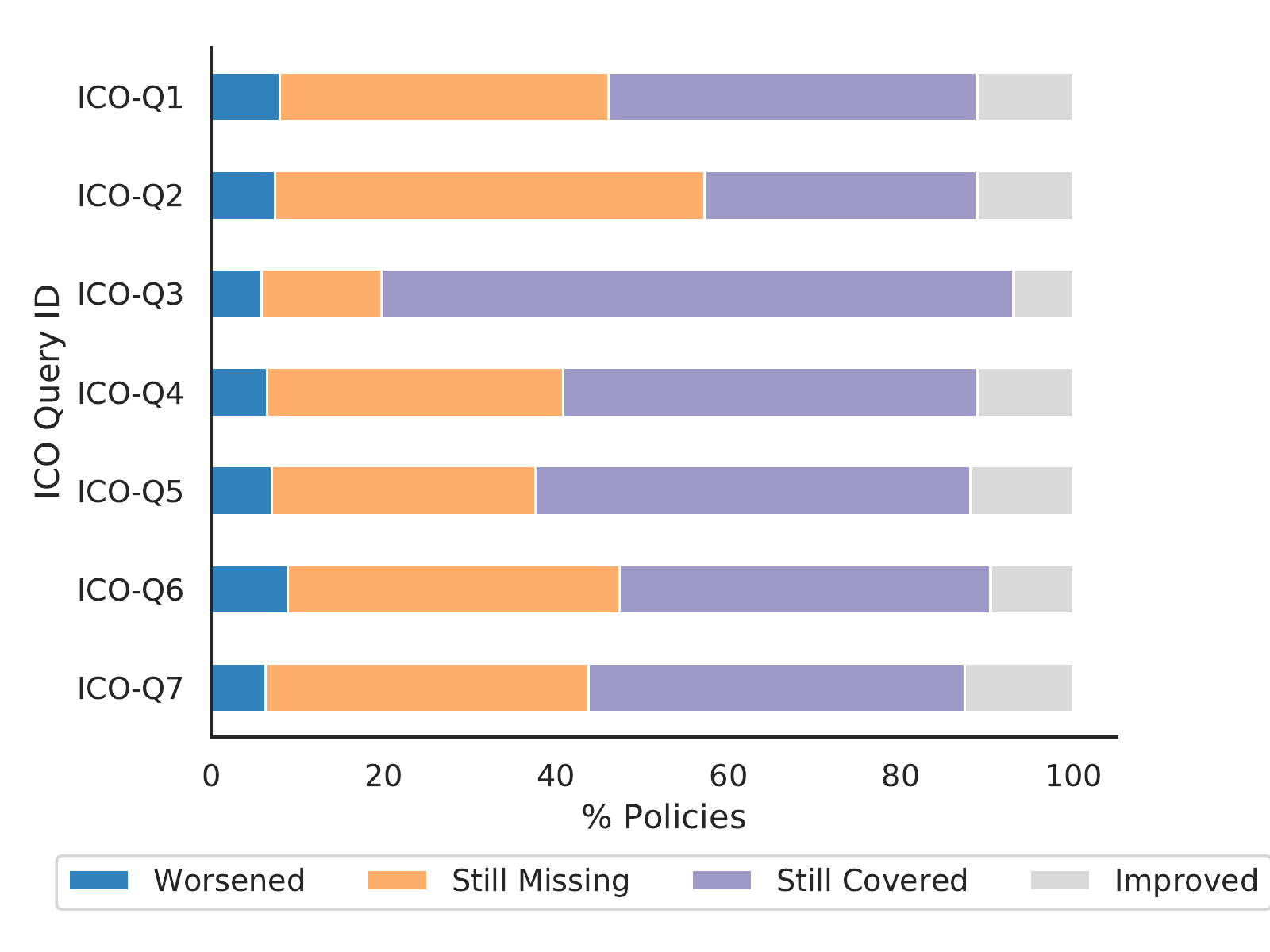}
    \caption{The comparison of ICO scores of \textit{Global} policies'
    \pregdpr and \postgdpr instances. The queries for the ICO checklist can be found in Table~\ref{tab:ico-qs}.}
    \label{fig:ico-total}
\end{figure}

Fig. \ref{fig:eu-ico-total} and \ref{fig:ico-total} show the percentage of \textit{EU} and \textit{Global} policies falling into each of the four cases for each ICO requirement. 
Similar to the methodology described in Sec.~\ref{sec:manual_analysis}, we computed the disagreement rate between queries built on automated labels of Polisis and those built on annotations of Subject Matter Experts (SMEs) from the OPP-115 dataset~\cite{Wilsonacl16}.
The disagreement rate averaged across 15 policies for Polisis-SMEs was found to be 0.21, which is comparable to the SME-SME disagreement rate of 0.19. We note that the disagreement rate is higher than that of coverage because of the added level of classification involved in the ICO queries.

\paragraph*{Case 1: Requirements Worsened}
For the vast majority of the policy sets, there has not been a noticeable decline in the covered ICO requirements. This decline has been limited to less than 12\% for both sets across all queries. We observe that ICO-Q6, referring to updating privacy policies if the site uses personal data for new purposes, exhibits the highest decline (11.7\%) among the seven requirements for both the \textit{EU} and \textit{Global} sets. An example of a policy with a worse score is Only Fans, whose \pregdpr policy contained the segment:
``{\small\textsf{We post any changes we make to the Policy on this page. If we make material changes to how we treat our Users’ personal information, we will notify you.}}'', had no similar statement in its \postgdpr instance.

\paragraph*{Case 2: Requirements Still Missing}
Despite the GDPR's emphasis on the right of the users to be informed, we find that many privacy policies are still lacking.
For instance, when it comes to specifying the sources of personal data obtained by first parties, there is still a large gap to fill. 
This is captured in ICO-Q2 for which we could not find evidence in 47.1\% of \textit{EU} policies and 49.8\% of \textit{Global}. Although companies are not always obtaining such data from third parties, these numbers are large enough to raise concerns about compliance with the ICO requirements. 

One notable exception is ICO-Q3, which is focused on specifying the third party entities \textbf{receiving} the data (only 13.0\% of \textit{EU} policies and 13.9\% of \textit{Global} policies maintained a score of 0 in their \pregdpr and \postgdpr versions).

\paragraph*{Case 3: Requirements Still Covered}
From Fig. \ref{fig:eu-ico-total} \& \ref{fig:ico-total}, we observe that except for ICO-Q2, at least 38\% of policies from both sets were, and are still compliant with the ICO requirements. This suggests that the policies are still being forthcoming with their practices, especially for ICO-Q3, 4, 5, and 7, which require the policy to talk about the source and recipients of collected data and users' rights to withdraw and access their data.

\paragraph*{Case 4: Requirements Improved}
Finally, a considerable portion of both policy sets, on average about 15\% for \textit{EU} and 10.7\% for \textit{Global}, have improved their coverage of the ICO requirements. ICO-Q7, informing users of their right to access their personal data, has seen the highest improvement (19.5\% for \textit{EU} and 12.6\% for \textit{Global}) among the policies in our datasets.
We observe that only a minority of the policies (mostly from the more popular websites) have started addressing the previously missing requirements (such as ICO-Q4, ICO-Q6). For example, NYTimes has added a new clause to address the requirement about notifying users about changes to the privacy policies:\\
``{\small\textsf{We evaluate this privacy policy periodically in light of changing business practices, technology, and legal requirements. As a result, it is updated from time to time. Any such changes will be posted on this page. If we make a significant or material change in the way we use or share your personal information, you will be notified via email and/or prominent notice within the NYT Services at least 30 days prior to the changes taking effect.}}''

\subsubsection*{Takeaways}
Similar to the coverage analysis, we find that the majority of the privacy policies, both inside and outside the EU, have incorporated the GDPR's privacy requirements we analyzed. On average, 59.3\% of the \textit{EU} \postgdpr policies and 58.2\% of the \textit{Global} \postgdpr policies meet our seven queries (combining Case 3 and Case 4). 
We also notice that policies state the recipients of user data (77.8\% \textit{EU}, 80.3\% \textit{Global}) and the collection sources of users' personal data (65\% \textit{EU}, 62.4\% \textit{Global}) fairly well across both sets of policies. 
On the other hand, both sets struggle in describing the categories of the collected data (41.3\% \textit{EU} and 42.8\% \textit{Global}); for our compliance analysis, only this single query had less than 50\% of the \postgdpr policies comply. 

Comparing the two sets, we observe that the average score of compliance is higher for the \textit{EU} policies. Also, the average margin of improvement (from \pregdpr to \postgdpr) across the chosen ICO requirements for the \textit{EU} set (4.8\%) is larger than that of the \textit{Global} set (3.5\%). 

\section{Specificity Analysis}
\label{sec:specificity}

\begin{table*}[t]
 \scriptsize
  \begin{center}
    \caption{Table of the specificity queries applied to each pre-GDPR and post-GDPR policy. Note the separate scoring Function for Q6 and Q7.}
    \label{tab:amb-qs}
    \begin{tabularx}{\textwidth}{m{2.5in} m{2.1in} m{2.3in}}
      \textbf{Description} & \textbf{Filtering Logic} & \textbf{Scoring Function}\\
      \midrule
      \textbf{Q1}: Quantify how specifically the policy indicates how the first party is obtaining user data.
      &
    Consider the set $S ={\{s_i\}}$ such that \par
    \makebox[2.7cm]{\category($s_i$)}=  \{\firsts\}
    \par
    \par
    \makebox[2.7cm]{\textit{action-first-party}($s_i$)}$\neq \phi$ 
      &
      Take $S_a \subset S$ such that \par
      \makebox[2.7cm]{\textit{action-first-party}($s_i$)}=  \unspec 
      \par
      The specificity score is: $1 - |S_a|/|S|$.
      \\
      \midrule 
      \textbf{Q2}: Quantify how specifically the policy indicates how the third party is collecting user data.
      &
        
    Consider the set $S ={\{s_i\}}$ such that \par
    \makebox[2.7cm]{\category($s_i$)}=  \{\thirds\}
    \par
    \makebox[2.7cm]{\textit{action-third-party}($s_i$)}$\neq \phi$
      &
      Take $S_a \subset S$ such that \par
      \makebox[2.7cm]{\textit{action-third-party}($s_i$)}=  \unspec 
      \par
      The specificity score is: $1 - |S_a|/|S|$.
      \\
      \midrule 
      \textbf{Q3}: Quantify how specifically the policy indicates the type of information accessed by the first party.
      &
        
    Consider the set $S ={\{s_i\}}$ such that \par
    \makebox[2.7cm]{\category($s_i$)}=  \{\firsts\}
    \par
        \par
        \makebox[2.7cm]{\textit{\pinfo}($s_i$)}$\neq \phi$
      &
      Take $S_a \subset S$ such that \par
        \makebox[2.7cm]{\textit{\pinfo}($s_i$)}=  \unspec
      \par
      The specificity score is: $1 - |S_a|/|S|$.
      \\
      \midrule 
      \textbf{Q4}: Quantify how specifically the policy indicates the type of information shared with the third party.
      &
        
    Consider the set $S ={\{s_i\}}$ such that \par
    \makebox[2.7cm]{\category($s_i$)}=  \{\thirds\}
    \par
        \makebox[2.7cm]{\textit{\pinfo}($s_i$)}$\neq \phi$
      &
      Take $S_a \subset S$ such that \par
        \makebox[2.7cm]{\textit{\pinfo}($s_i$)}=  \unspec
      \par
      The specificity score is: $1 - |S_a|/|S|$.
      \\
      \midrule
      \textbf{Q5}: Quantify how specifically the policy indicates how the third party is receiving user information.
      &
        
    Consider the set $S ={\{s_i\}}$ such that \par
    \makebox[2.7cm]{\category($s_i$)}=  \{\thirds\}
      &
      Take $S_a \subset S$ such that \par
        \makebox[2.7cm]{\textit{third-party-entity}($s_i$)}=  \unspec
      \par
      The specificity score is: $1 - |S_a|/|S|$.
      \\
      \midrule 
      \textbf{Q6}: Quantify how specifically the policy covers first party collection purposes relative to all possible purposes in our taxonomy.
      &
    Let $P$ be the set of all purposes. \par
    Let $P_s$ be the set of all purposes $p$  \par
    such that $\exists$ a segment $s_i$ with: \par
    \makebox[2.7cm]{\category($s_i$)} =  \{\firsts\}
    \par
    \makebox[2.7cm]{$p$} $\in$ \textit{purpose}($s_i$)
      &
     The specificity score is $|P_s|/|P|$. 
      \\
      \midrule 
      \textbf{Q7}:  Quantify how specifically the policy covers third party sharing purposes relative to all possible purposes in our taxonomy. 
      &
    Let $P$ be the set of all purposes.\par
    Let $P_s$ be the set of all purposes $p$  \par
    such that $\exists$ a segment $s_i$ with: \par
    \makebox[2.7cm]{\category($s_i$)}=  \{\thirds\}
    \par
    \makebox[2.7cm]{$p$} $\in$ \textit{purpose}($s_i$)
      &
      The specificity score is $|P_s|/|P|$. 
      \\
      \midrule 
      \textbf{Q8}:  Quantify how specifically the policy indicates the purpose for data retention.
      &
    Consider the set $S ={\{s_i\}}$ such that \par
    \makebox[2.7cm]{\category($s_i$)}=  \{\retention\}
    \par
    \makebox[2.7cm]{\textit{purpose}($s_i$)} $\neq \phi$
      &
      Take $S_a \subset S$ such that \par
      \makebox[2.7cm]{\textit{purpose}($s_i$)}=  \unspec 
      \par
      The specificity score is: $1 - |S_a|/|S|$.
      \\
    \end{tabularx}
  \end{center}
\end{table*}

Compliance and coverage describe whether a policy mentions a privacy practice. Merely mentioning a privacy practice, however, does not fully satisfy transparency; it is not clear whether these practices are covered in general or specific terms. We quantify this aspect through specificity queries. For example, the statement \textsf{``We collect your personal information \ldots''} covers collection by the first party but is not specific as to which type of personal information is being collected; a specific statement would be \textsf{``We collect your health data \ldots''}
In this section, we aim to assess the change in the level of specificity present in the privacy policies.

We use the filtering-scoring approach of Sec.~\ref{sec:reqs} to quantify the policy's specificity of describing a privacy practice. Table~\ref{tab:amb-qs} describes the eight specificity queries (Q1 $\rightarrow$ Q8) that quantify how explicit the privacy policy is in describing: how first party is collecting data, how third-party is obtaining data, the information type collected, information type shared, purposes for data collection, purposes for data sharing, and purposes for data retention. For all the queries, a higher score indicates higher specificity.

The reader might notice a discrepancy in Table~\ref{tab:amb-qs} in the scoring step for queries focusing on the \textit{purpose} attribute (first party (Q6) and third party (Q7)) vs. the rest of the queries.
We treated these cases differently due to the way Polisis interprets privacy policies.
Within the Polisis system, $\purpose(s)=\unspec$ does not always imply a missing purpose of data collection/sharing. Instead, it might indicate that the purpose is not the subject of that segment.
Hence, most of the segments that focus on the data types being collected or shared will carry an \unspec purpose label. 
Accordingly, we quantify purpose specificity, in the first party and third party contexts, as the ratio of the number of stated purposes in the policy ($|P_s|$) to the total number of possible purposes ($|P|$). 
On the other hand data retention is typically addressed by one or two segments in the policy; the segments almost always describe the purpose. If those segments have $\purpose(s)=\unspec$, then we expect that the policy is not being specific in explaining the purpose for data retention.

\begin{figure}[t]
    \centering
  \includegraphics[width=1\columnwidth]{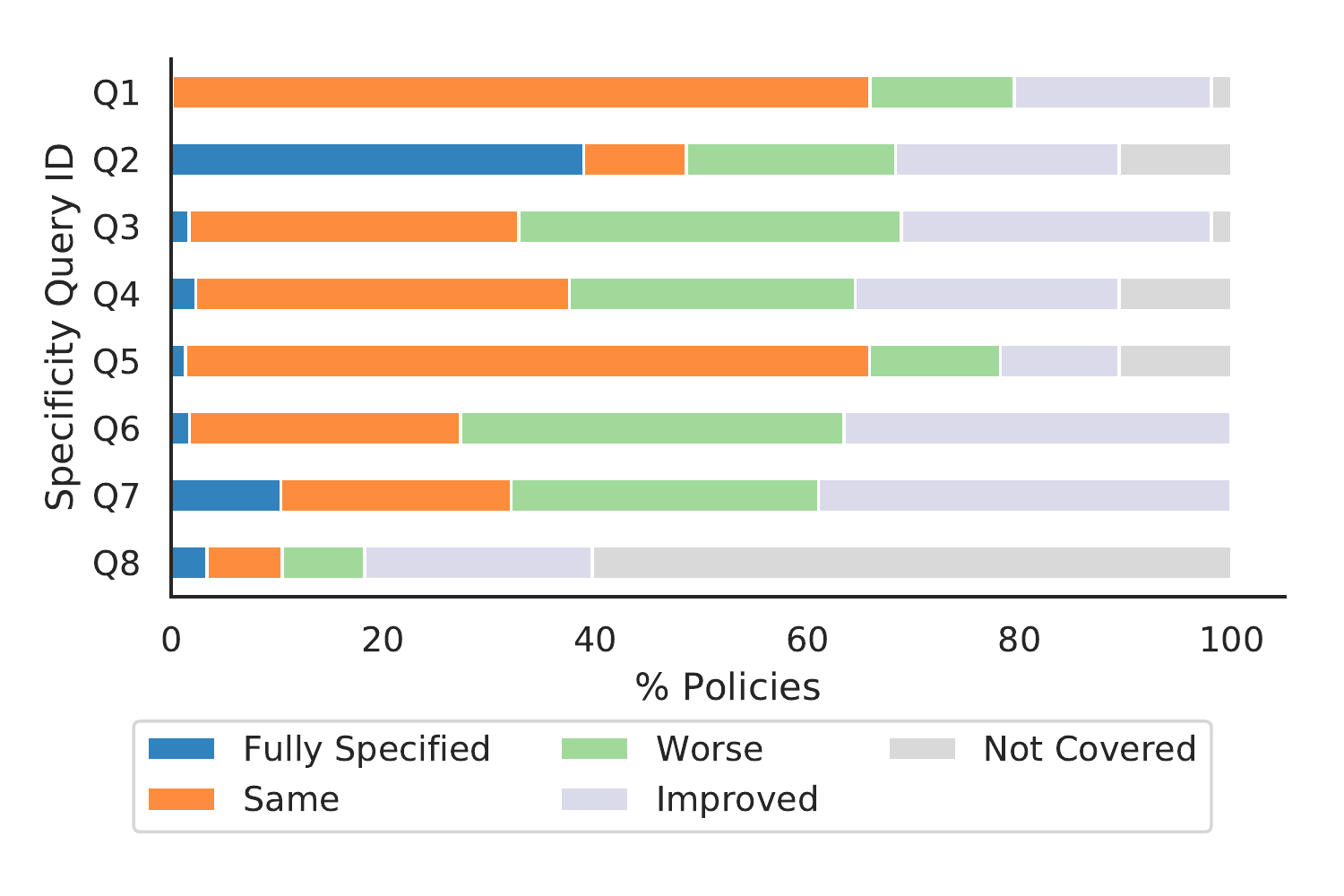}
    \caption{ The comparison specificity scores of \textit{EU} policies' \pregdpr and \postgdpr instances.}
    \label{fig:amb-total_eu}
\end{figure}

\begin{figure}[t]
    \centering
  \includegraphics[width=1\columnwidth]{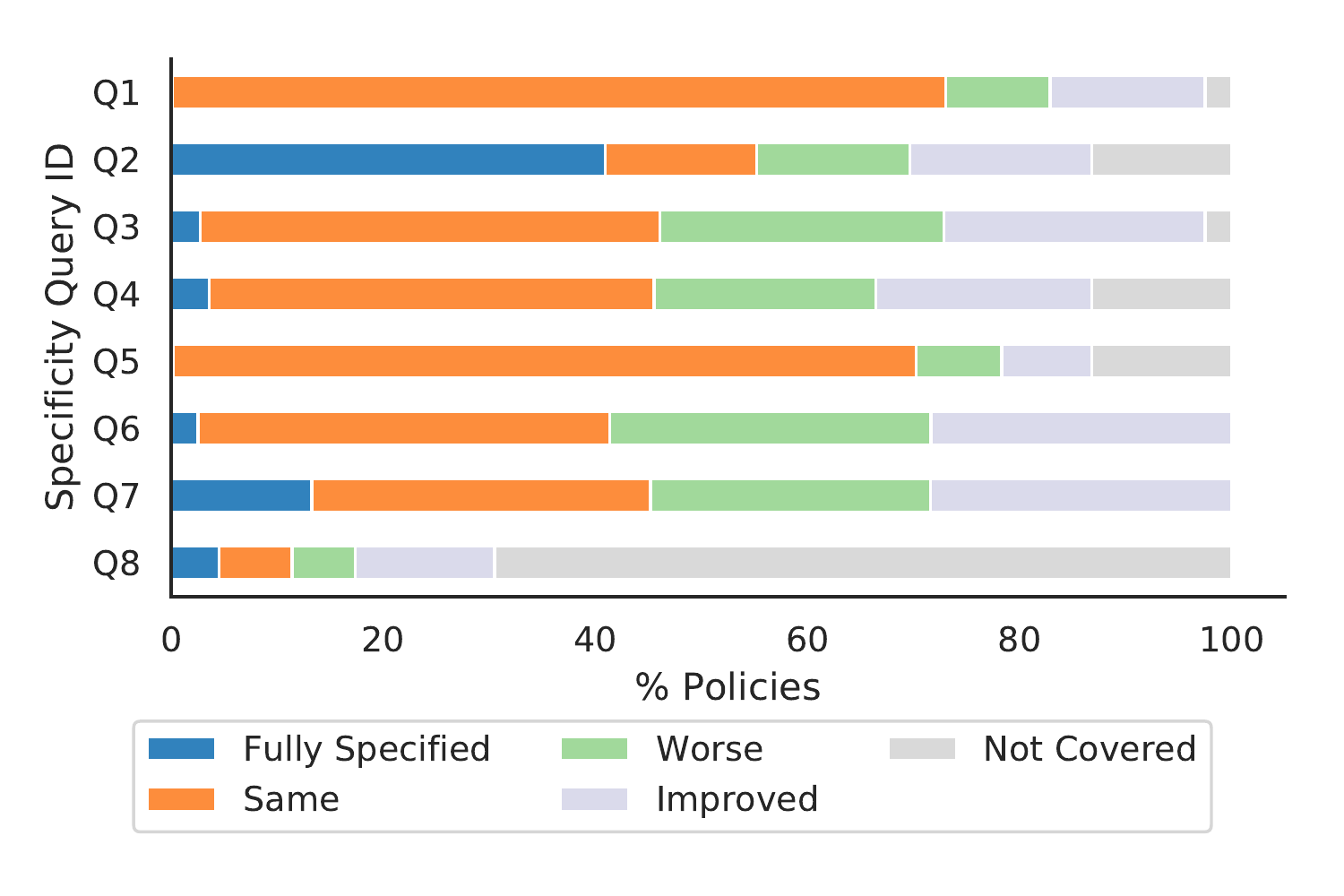}
    \caption{ The comparison specificity scores of \textit{Global} policies' \pregdpr and \postgdpr instances.}
    \label{fig:amb-total_global}
\end{figure}

 We analyze the evolution of the eight specificity scores between each policy's \pregdpr and \postgdpr instances. We also note that manual verification for specificity queries is not suitable here because segments in Polisis are different than the segments in the OPP-115 dataset, and the specificity scores depend heavily on the number of segments with a particular label. We consider four cases in Fig.~\ref{fig:amb-total_eu} and Fig.~\ref{fig:amb-total_global}:

 \begin{itemize}
    \item \textbf{Query Not Covered:} A policy did not cover the requirement in both its \pregdpr and \postgdpr versions; i.e. $|S| = 0$ or $|P|=0$ for both the versions.
    \item \textbf{Same Specificity:} A policy maintains the same specificity level about a practice between the two versions, but the score is not equal to 1.
    \item \textbf{Fully Specified:} A policy continues to be fully specific about a practice; i.e. the score $= 1$ for both the versions. 
    \item \textbf{Worse Specificity:} The \postgdpr version of the policy has a lower specificity score than its \pregdpr instance counterpart.
    \item \textbf{Improved Specificity:} The \postgdpr version of the policy has a higher specificity score than its \pregdpr instance counterpart.
\end{itemize}

\paragraph*{Case 1: Query Not Covered}
 Consistent with the results of Sec.~\ref{sec:cat_coverage}, we find that the purpose of data retention practice (Q8) is not frequently covered among both sets of studied policies. Also, we find that approximately 10.6\% of \textit{EU} policies and 13.2\% of \textit{Global} policies do not cover the queries Q2, Q4, and Q5 because they do not cover the third party sharing category. This observation is relatively consistent with the results of Sec.~\ref{sec:cat_coverage} where around 20\% of the \pregdpr and \postgdpr instances of policies do not cover this category.

\paragraph*{Case 2: Same specificity}
A large portion of the policies in both sets exhibited the same specificity scores for the analyzed privacy practices, particularly for queries Q1 and Q5. This result is not surprising given that about 20\% of \textit{EU} policies and 35\% of \textit{Global} policies did not change between \pregdpr and \postgdpr instance, as seen in Fig.~\ref{fig:fuzzy_ratio_combined}. For the other policies, they maintain the same specificity levels even when their content changes.

\paragraph*{Case 3: Fully Specified}  
  For the privacy practices covered in Q2, the specificity values stay at one for 40\% (and to a lower degree Q7) of the studied policies for both datasets. These subsets of policies mention the specific methods of collecting user and sharing data. We also observe that the portion of fully specified policies is fairly consistent between the \textit{EU} and \textit{Global} sets for all queries.

\paragraph*{Case 4: Worse Specificity}
 Interestingly, we observe a considerable portion of policies exhibiting lower specificity in describing their privacy practices. We attribute this reason to the policies trying to be more comprehensive and general in describing the data practices at the expense of the specificity of the data practices clauses. 
 This is a representative example from the \postgdpr snapshot of hrc.org\footnote{https://www.hrc.org/hrc-story/privacy-policy}: \\
{\small\textsf{``We also may be required to release information if required to do so by law or if, in our sole discretion, such disclosure is reasonably necessary to comply with legal process or to protect the rights, property, or personal safety of our web site, HRC, our officers, directors, employees, members, other users, and/or the public.  We may also share personal information in the event of an organizational restructuring''.}}
While the \pregdpr snapshot contained segments related to sharing data with third-party entities, this newly added segment does not specify the type of personal information released.

\paragraph*{Case 5: Improved Specificity}
Finally, we observe that a large portion of the privacy policies have improved their specificity by using more precise phrases to describe the data collected and shared along with the purposes. For both sets, five of the eight queries had a higher number of policies with improving specificity than the number of policies with reduced specificity. This event occurred for queries Q1, 2, 7, and 8 for each set. Additionally, the \textit{EU} set had more positive changes than negative changes for query Q6 (around first party purposes), and the \textit{Global} set had more for query Q5 (around third party entities). 
Here is a representative example of improved specificity for Q1 (around data collection method) from legacy.com\footnote{https://www.legacy.com/about/privacy-policy}. 
\newline
\pregdpr: {\small\textsf{``Examples of Personally Identifiable Information we may collect include name, postal address, email address, credit card number and related information, and phone number....}}''\newline
\postgdpr: {\small\textsf{``...the personal data you provide \textit{when you register to use our Services}, including your name, email address, and password; the personal data that may be contained in any video, photo, image...''}}
We note here that \pregdpr version did not mention how the personal data was being collected, while the \postgdpr version specifically mentions it as when the user registers to use the service.

\subsubsection*{Takeaways}
In conclusion, privacy policies appear to be more specific post the GDPR. While many of them have maintained the same specificity levels (due to unchanged policies or low coverage of data practices), a considerable number of policies has been changed. Of those policies, for both sets, a minority encountered reduced specificity according to our metrics; to comply with the GDPR, they have tried to be more comprehensive in describing their practices at the expense of being less specific. The majority of the policies that changed, however, have been more specific in informing the users about their privacy practices. 

When comparing the two sets, the \textit{EU} set exhibits a higher portion of improving policies than the \textit{Global} set for all the specificity queries. The \textit{EU} set also had a higher portion of policies with lower scores for each of the eight specificity queries. Recall that the \textit{EU} dataset has higher coverage of the related privacy practices (Sec.~\ref{sec:cat_coverage}). This result suggests that simply modifying a policy to increase coverage does not guarantee transparency to the data subjects. This can also be seen by analyzing policies whose coverage improved for a particular category, \textit{e.g.} data retention. We find that more than 58\% of policies with improved data retention coverage are not fully specific (\textit{i.e.} they have a specificity score of less than 1) suggesting that, despite better coverage, the transparent disclosure of the privacy information  is yet to be fully achieved.

\section{Limitations}
\label{sec:limitations}
Despite our efforts to cover a diverse set of websites and to understand privacy policies' evolution from multiple angles, we acknowledge that this work has several limitations.

First, our approach in assessing the content of privacy policies does not fully capture all the efforts introduced due to the GDPR. For instance, the concept of layered privacy notices has been adopted by several companies to give users two levels of understanding: an overview of high-level practices and an in-depth description of these practices. Unfortunately, this is difficult to automatically analyze as it can come in a variety of formats, such as multi-page policies.

Second, our study is limited to the English language policies as we wanted to leverage the existing techniques of advanced semantic understanding of natural language. We made this trade-off for depth vs. language-coverage and decided not to use a keyword-based analysis.

Third, the use of automated approaches for selecting privacy policies and analyzing them is inherently error-prone. Hence, our specificity and compliance analysis might have been affected by the errors made with the machine learning models. This is accentuated by the fact that privacy policies are complex and even human annotators disagree on the interpretations sometimes, as we found in our manual verification. Nevertheless, with the recent success and increasing accuracy levels across different tasks~\cite{Wilsonacl16, harkous2018polisis}, we believe that such techniques, coupled with manual post-analysis, are a highly effective venue for in-depth analysis at scale.

Fourth, we restricted our user study to focus on the appearance of privacy policies, rather than their content. A more comprehensive user study, as a dedicated research project, is needed to gain a deeper understanding of how much the readability of the policies has evolved.

\section{Related Work}
\label{sec:related}

In the following, we survey the evolution of the privacy policies' landscape, particularly in relation to regulatory intervention. We also describe the recent research that studies the GDPR's impact on privacy policies.

\subsubsection*{Evolution of the Privacy Policies Landscape}
In 2002, the Progress and Freedom Foundation (PFF) studied a random sample of the most visited websites and found that, compared to two years earlier, websites were collecting less personally identifiable information and offering more opt-in choices and less opt-out choices~\cite{adkinson2002privacy}. 
Another longitudinal analysis has been performed by Milne and Culnan in the 1998-2001 period, confirming the positive change in the number of websites including notices about information collection, third-party disclosures, and user choices~\cite{milne2002using}. 
In the same period, Liu and Arnett found that slightly more than 30\% of Global 500 Web sites provide privacy policies on their home page~\cite{liu2002raising}. 

Despite the increased proliferation of privacy policies, their lack of clarity was one of the primary motivations of the regulatory measures before the GDPR. In 2004, Antón et al. showed that 40 online privacy statements from 9 financial institutions have questionable compliance with the requirements of the Gramm-Leach-Bliley Act (GLBA). They assessed the requirements of having ``clear and conspicuous'' policies via keyword-based investigation and readability metrics~\cite{anton2004financial}. 
In 2007, Antón et al. studied the effect of the Health Information and Portability Accountability Act (HIPAA) via a longitudinal study of 24
healthcare privacy policy documents from 9 healthcare Web sites. A similar conclusion held: although HIPAA has resulted in more
descriptive policies, the overall trend was reduced readability and less clarity. 
Resorting to a user study, in 2008, Vail et al. showed that users perceive traditional privacy policies (in paragraph-form) to be more secure than shorter, simplified alternatives. However, they also demonstrated that these policies are significantly more difficult to comprehend than other formats~\cite{vail2008empirical}. 

In a recent study, Turow et al. studied the surveys around the ``privacy policy'' as a label between 2003 and 2015 in the US. They found that the users' misplaced confidence in this label, not only carries implication on their personal lives but affects their actions as citizens in response to government regulations or corporate activities~\cite{turow2018persistent}.

\subsubsection*{Privacy Policies After the GDPR}
Since the GDPR had been enforced on May 25, 2018, a few studies have investigated its impact on privacy practices of companies. Despite the initial trials with automated approaches in these studies, they have been limited in terms of scale, which is the main goal behind automation. Contissa et al. conducted a preliminary survey of 14 privacy policies of the top companies as an attempt to measure the compliance of these companies with the GDPR automatically. They found a frequent presence of unclear language, problematic processing, and insufficient information. They used Claudette, a recent system designed for the detection of such types of issues in privacy policies~\cite{lippi2018claudette}. Tesfay et al. introduced a tool, inspired by the GDPR to classify privacy policy content into eleven privacy aspects~\cite{tesfay2018privacyguide}. They validated their approach with ten privacy policies. 

The first large-scale study concurrent to ours is that by Degeling et al. who performed a longitudinal analysis of the privacy policies and cookie consent notices of 6,759 websites representing the 500 most popular websites in each of the 28 member states of the EU~\cite{degeling2018we}. They found that the number of websites with privacy policies has increased by 4.9\% and that 50\% of websites updated
their privacy policies just before the GDPR came into action in May 2018. Unlike our work, however, their work has been focused on cookie consent notices and terminology-based analysis of privacy policies, without an in-depth tackling of the semantic change of privacy policies.

\section{Takeaways and Conclusions}
\label{sec:conclusion}

In this paper, we seek to answer a question about the impact of the recently introduced General Data Protection Regulation on website privacy policies. To answer this question, we analyze a sample of 6,278 unique English-language privacy policies from inside and outside the EU. Using the \pregdpr and \postgdpr of each policy we study the changes along five dimensions: presentation, textual features, coverage, compliance, and specificity.

The results of our tests and analyses suggest that the GDPR has been a catalyst for a major overhaul of the privacy policies inside and outside the EU. This overhaul of the policies, manifesting in extensive textual changes, especially for the EU-based websites, does not necessarily come at a benefit to the users. Policies have become considerably longer (a fifth longer in the \textit{Global} set and a third longer in the \textit{EU} set). Our presentation analysis, however, identified a positive trend in user experience, specifically for the \textit{EU} policies; such a trend was not found for \textit{Global} policies. 

We discovered another positive development in the high-level privacy category coverage of policies. For both \textit{EU} and \textit{Global} sets, the portion of policies mentioning these topics increased, especially in categories that were previously underrepresented. Interestingly, once more, we observed a more significant upward trend from the \textit{EU} policies, and this disparity between the two sets continued in our compliance analysis. The average portion of compliant  \textit{EU} policies was larger than that of the \textit{Global} policies (according to our queries). While the majority of policies in both sets scored full marks for each compliance query, the analysis revealed inconsistency between coverage and compliance. Simply covering a high-level privacy topic does not ensure the level of detail required by the GDPR is met. This discrepancy emerges in our specificity analysis. Again the margin of improvement was larger for \textit{EU} policies compared to \textit{Global}. However, we observed that in several cases the improvement in coverage resulted in reduced specificity. 

In summary, we observe two major trends from our analysis. First, while the GDPR has made a positive impact on the overall incorporation of privacy rights and information, its influence is more pronounced inside the EU, its primary target region. Second, although policies are becoming consistently longer and covering more privacy topics, more change is to come before we reach a stable status of full-disclosure and transparency.

\bibliographystyle{IEEEtranS}
\bibliography{IEEEabrv,z_0_refs.bib}

\appendix
\clearpage

\section{Policy Classifier Architecture}
\label{sec:pol_classifier}

\begin{figure}[h!]
    \centering
    \includegraphics[width=\linewidth]{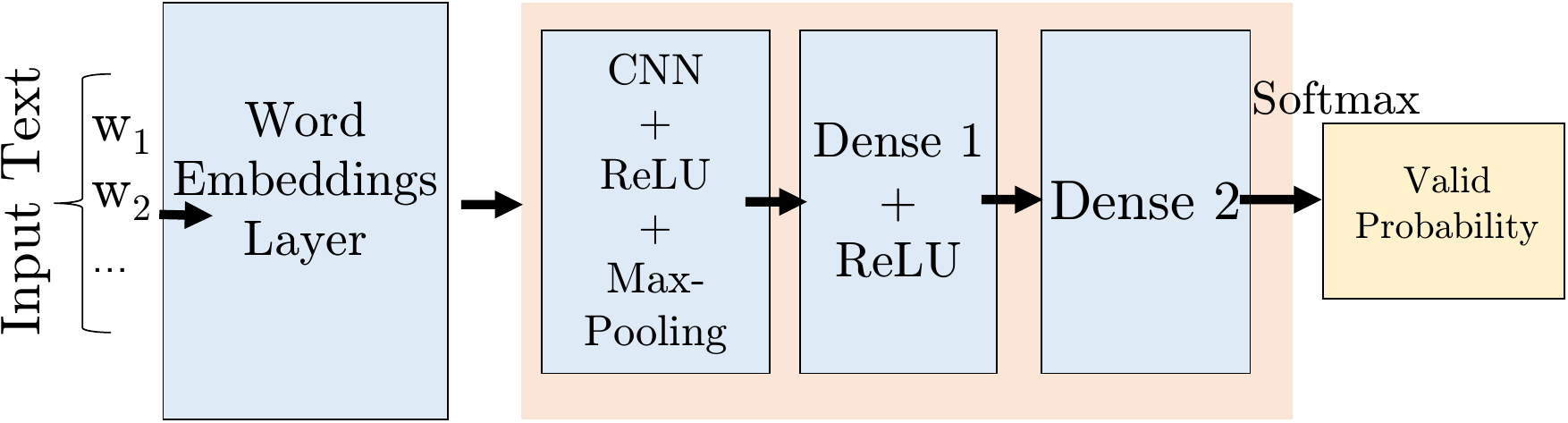}
    \caption{Architecture of the \textit{``Is Policy?''} classifier used to determine whether a privacy policy is valid. Hyperparameters used: Embeddings size: 100, Number of filters: 250, Filter Size: 3, Dense Layer Size: 250, Batch Size: 32}
    \label{fig:classifier_policy}
\end{figure}

Fig.~\ref{fig:classifier_policy} shows the detailed architecture of the single-label classifier used in Sec.~\ref{sec:methodology} for checking whether the crawled pages are valid privacy policies.
The input text, obtained from the web page, is tokenized into words, using the Penn Treebank Tokenizer in \textit{nltk}\footnote{\url{http://www.nltk.org/}}.
Then the words are mapped into vectors at the word embeddings layer. The word vectors are input to a convolutional layer, followed by a Rectified Linear Unit (ReLU) and a Max-pooling layer. The next layer is a fully connected (dense) layer followed by another ReLU. Finally, we apply a softmax on the output dense layer to obtain a probability distribution over the two labels ``valid'' and ``invalid''. For more details about this kind of classifiers we refer the reader to the work of Kim~\cite{Kim14} on sentence classification.

The data used to train the classifier was composed of (1) a set of 1,000 privacy policies labeled as \textit{valid} from the ACL/COLING 2014 privacy policies' dataset released by Ramanath et al.~\cite{RamanathLSS14} and (2) an \textit{invalid} set consisting of the text from 1,000 web pages, fetched from random links within the homepages of the top 500 Alexa websites. We ensured that the latter pages do not have any of the keywords associated with privacy policies in their URL or title. The data was split into an 80\% training set and a 20\% testing set, and the classifier yielded a 99.09\% accuracy on the testing set.

\section{User Survey}

\begin{figure}[ht!]
    \centering
    \includegraphics[width=\linewidth]{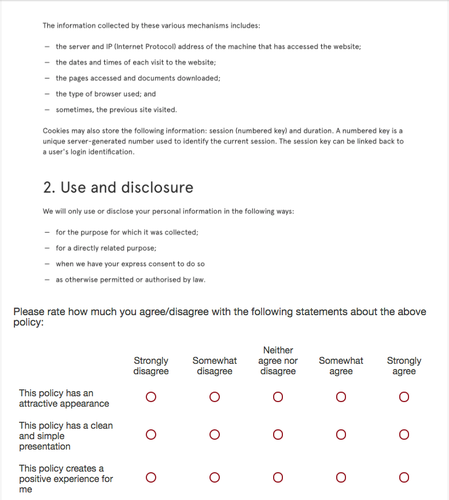}
    \caption{Example step from our user survey where the users had to respond to the three questions}
    \label{fig:survey_snap}
\end{figure}

In Fig.~\ref{fig:survey_snap}, we show a screenshot of an example step from the user survey that we presented to the users in Sec.~\ref{sec:presentation}.
\label{sec:appndx_survey}

\end{document}